\documentclass[iop,apj]{emulateapj}
\newcommand\pasa{PASA}%
\submitted{ApJ, in press}

\defcitealias{ngs+08}{N08}

\newcommand{\kms}{\,km\,s$^{-1}$}

\newcommand{\mjb}{\,mJy\,beam$^{-1}$}

\begin{document}
\title{EVOLUTION OF THE RADIO REMNANT OF SUPERNOVA 1987A: MORPHOLOGICAL CHANGES
FROM DAY 7000}

\author{C.-Y. Ng\altaffilmark{1},
G. Zanardo\altaffilmark{2},
T. M. Potter\altaffilmark{2},
L. Staveley-Smith\altaffilmark{2,3},
B. M. Gaensler\altaffilmark{3,4},
R. N. Manchester\altaffilmark{5},
and
A.~K.~Tzioumis\altaffilmark{5} } 

\altaffiltext{1}{Department of Physics, The University of Hong Kong, Pokfulam Road, Hong Kong}
\altaffiltext{2}{International Centre for Radio Astronomy Research (ICRAR) --
The University of Western Australia, Crawley, WA 6009, Australia}
\altaffiltext{3}{Australian Research Council, Centre of Excellence for All-sky
Astrophysics (CAASTRO)}
\altaffiltext{4}{Sydney Institute for Astronomy, School of Physics, The
University of Sydney, NSW 2006, Australia}
\altaffiltext{5}{CSIRO Astronomy and Space Science, Australia Telescope
National Facility, Marsfield, NSW 1710, Australia}

\email{ncy@bohr.physics.hku.hk}

\begin{abstract}
We present radio imaging observations of supernova remnant 1987A at 9\,GHz,
taken with the Australia Telescope Compact Array over 21 years from 1992 to
2013. By employing a Fourier modeling technique to fit the visibility data, we
show that the remnant structure has evolved significantly since day 7000
(mid-2006): the emission latitude has gradually decreased, such that the
overall geometry has become more similar to a ring structure. Around the same
time, we find a decreasing trend in the east-west asymmetry of the surface
emissivity. These results could reflect the increasing interaction of the
forward shock with material around the circumstellar ring, and the relative
weakening of the interaction with the lower-density material at higher
latitudes. The morphological evolution caused an apparent break in the remnant
expansion measured with a torus model, from a velocity of $4600^{+150}_
{-200}$\kms\ between day 4000 and 7000 to $2400^{+100}_{-200}$\kms\ after day
7000. However, we emphasize that there is no conclusive evidence for a
physical slowing of the shock at any given latitude in the expanding remnant,
and that a change of radio morphology alone appears to dominate the evolution.
This is supported by our ring-only fits which show a constant expansion of
$3890\pm50$\kms\ without deceleration between days 4000 and 9000. We suggest
that once the emission latitude no longer decreases, the expansion velocity
obtained from the torus model should return to the same value as that measured
with the ring model.

\end{abstract}

\shorttitle{Evolution of the Radio Remnant of SN 1987A}
\shortauthors{Ng et al.}


\section{Introduction}
\bibpunct[ ]{(}{)}{,}{a}{}{;} 
The remarkable supernova \object[SN 1987A]{(SN) 1987A} in the Large Magellanic
Cloud has enabled detailed studies of many fields in astrophysics, from
massive star evolution to the SN explosion mechanism to the earliest
stage of supernova remnants \citep[SNRs; see reviews in][]{iwm07}. As well as
being the brightest SN over the past 400\,yr, SN 1987A was a highly
unusual event. In particular, the progenitor was surrounded by a peculiar
triple-ring nebula \citep{bkh+95}, which could have resulted from a binary
merger of the progenitor 20000\,yr prior to the explosion \citep{mp07}. Over
the past decade, the remnant has undergone a major evolution since the
shock collision with the inner equatorial ring, resulting in rapid brightening
of the radio and soft X-ray emission \citep[see][and references therein]{zsb+10,hbd+13}.

\begin{deluxetable*}{lcclcc}
\tablecaption{Observational Parameters for the Datasets Used in This
Study\label{table:obs}}
\tablewidth{0pt}
\tabletypesize{\small}
\tablehead{
\colhead{Observing Date} & \colhead{Days since} & \colhead{Array} &
\colhead{Center Frequency\tablenotemark{a}} & \colhead{Time on} &
\colhead{Epoch Shown}  \\ \colhead{} & \colhead{Supernova} &
\colhead{Configuration} & \colhead{(MHz)} & \colhead{Source (hr)} &
\colhead{in Figure~\ref{fig:super}\tablenotemark{b}}}
\startdata
1992 Jan 14& 1786 & 6B & 8640  & 12 & \nodata \\
1992 Mar 20& 1852 & 6A & 8640  & 10 & \nodata \\
1992 Oct 21& 2067 & 6C & 8640, 8900 & 13 & 1992.9 \\
1993 Jan 4 & 2142 & 6A & 8640, 8900 & 9 & 1992.9 \\
1993 Jan 5 & 2143 & 6A & 8640, 8900 & 6 & 1992.9 \\
1993 Jun 24& 2313 & 6C & 8640, 8900 & 8 & 1993.6 \\
1993 Jul 1 & 2320 & 6C & 8640, 8900 & 10 & 1993.6 \\
1993 Oct 15& 2426 & 6A & 8640, 9024 & 17 & 1993.6 \\
1994 Feb 16& 2550 & 6B & 8640, 9024 & 9 & 1994.4 \\
1994 Jun 27-28& 2681 & 6C & 8640, 9024 & 21 & 1994.4 \\
1994 Jul 1 & 2685 & 6A & 8640, 9024 & 10 & 1994.4 \\
1995 Jul 24& 3073 & 6C & 8640, 9024 & 12 & 1995.7 \\
1995 Aug 29& 3109 & 6D & 8896, 9152 & 7 & 1995.7 \\
1995 Nov 6 & 3178 & 6A & 8640, 9024 & 9 & 1995.7 \\
1996 Jul 21& 3436 & 6C & 8640, 9024 & 14 & 1996.7 \\
1996 Sep 8 & 3485 & 6B & 8640, 9024 & 13 & 1996.7 \\
1996 Oct 5 & 3512 & 6A & 8896, 9152 & 8  & 1996.7 \\
1997 Nov 11& 3914 & 6C & 8512, 8896 & 7  & 1998.0 \\
1998 Feb 18& 4013 & 6A & 8896, 9152 & 10 & 1998.0  \\
1998 Feb 21& 4016 & 6B & 8512, 9024 & 7  & 1998.0 \\
1998 Sep 13& 4220 & 6A & 8896, 9152 & 12 & 1998.9  \\
1998 Oct 31& 4268 & 6D & 8502, 9024 & 11 & 1998.9  \\
1999 Feb 12& 4372 & 6C & 8512, 8896 & 10 & 1999.7 \\
1999 Sep 5 & 4577 & 6D & 8768, 9152 & 11 & 1999.7 \\
1999 Sep 12& 4584 & 6A & 8512, 8896 & 14 & 1999.7 \\
2000 Sep 28& 4966 & 6A & 8512, 8896 & 10 & 2000.9 \\
2000 Nov 12& 5011 & 6C & 8512, 8896 & 11 & 2000.9 \\
2001 Nov 23& 5387 & 6D & 8768, 9152 & 8 & 2001.9 \\
2002 Nov 19& 5748 & 6A & 8512, 8896 & 8 & 2003.0 \\
2003 Jan 20& 5810 & 6B & 8512, 9024 & 9 & 2003.0 \\
2003 Aug 1 & 6003 & 6D & 8768, 9152 & 10 & 2003.6 \\
2003 Dec 5 & 6129 & 6A & 8512, 8896 & 9 & 2004.0 \\
2004 Jan 15& 6170 & 6A & 8512, 8896 & 9 & 2004.0 \\
2004 May 7 & 6283 & 6C & 8512, 8896 & 9 & 2004.4 \\
2005 Mar 25& 6605 & 6A & 8512, 8896 & 9 & 2005.2 \\
2005 Jun 21& 6693 & 6B & 8512, 8896 & 9 & 2005.5 \\
2006 Mar 28& 6973 & 6C & 8512, 8896 & 9 & 2006.2 \\
2006 Jul 18& 7085 & 6A & 8512, 8896 & 9 & 2006.5 \\
2006 Dec 8 & 7228 & 6B & 8512, 9024 & 8 & 2006.9 \\
2008 Jan 4 & 7620 & 6A & 8512, 9024 & 11 & 2008.0 \\ 
2008 Apr 23& 7730 & 6A & 8512, 8896 & 11 & 2008.3 \\
2008 Oct 11& 7901 & 6A & 8512, 8896 & 11 & 2008.8 \\
2009 Jun 6 & 8139 & 6A & 9000  & 11 &     2009.4 \\
2010 Jan 23& 8370 & 6A & 9000  & 11 &     2010.1 \\
2010 Apr 11& 8448 & 6A & 9000  & 11 &     2010.3 \\
2011 Jan 25& 8737 & 6A & 9000  & 11 &     2011.1 \\
2011 Apr 22& 8824 & 6A & 9000  & 11 &     2011.3 \\
2012 Jan 12& 9089 & 6A & 9000  & 11 &     2012.0 \\
2012 Jun 5 & 9233 & 6D & 9000  & 11 &     2012.4 \\
2012 Sep 1 & 9321 & 6A & 9000  & 10 &     2012.7 \\
2013 Mar 7 & 9509 & 6A & 9000  & 11 &     2013.2 \\
2013 May 5 & 9568 & 6C & 9000  & 11 &     2013.3
\enddata
\tablenotetext{a}{Since the CABB upgrade in mid-2009, data have been recorded
over a 2-GHz bandwidth. However, in this analysis we used two 104-MHz subbands
with center frequencies of 8.512\,GHz and 8.896\,GHz, for a consistency with
the bandwidth of pre-CABB data.}
\tablenotetext{b}{Some datasets have been averaged together to generate the
corresponding images in Figure~\ref{fig:super} for the listed epoch.}
\end{deluxetable*}

\bibpunct[, ]{(}{)}{,}{a}{}{;} 
Radio emission of SNR 1987A is believed to be non-thermal synchrotron
radiation emitted by energetic particles accelerated in shocks. Since the
remnant emerged in mid-1990 \citep{tcm+90,smk+92}, it has been monitored
regularly at different frequencies using the Australia Telescope Compact Array
\citep[ATCA; see][and references therein]{sgm+07, zsb+10}. The flux density
was found to have increased exponentially from day 5000 to 8000 with a
progressively flatter spectrum \citep{zsb+10}, indicating increasingly
efficient particle acceleration processes.

\begin{figure*}[!ht]
\epsscale{1}
\vspace*{-5mm}
\plotone{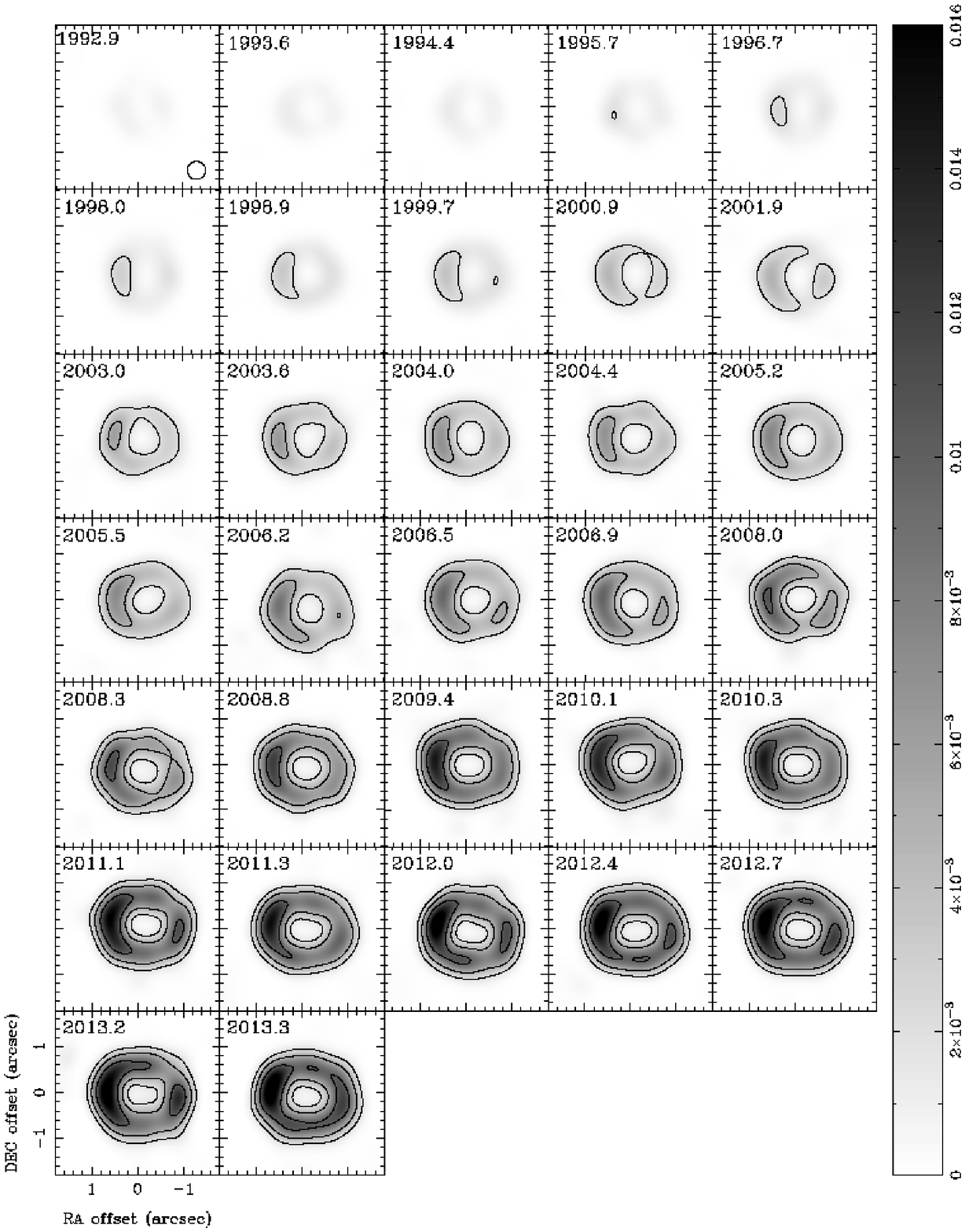}
\caption{Super-resolved 9\,GHz ATCA images of SN 1987A over the period
1992--2013. Some early epochs have been averaged to boost the signal-to-noise
ratio (see Table~\ref{table:obs}). The gray scale is linear ranging from 0 to
16\mjb\ and the contours are at levels of 2, 5, 10, and 15\mjb. The
synthesized beam, which has an FWHM of 0\farcs4, is shown in the first panel.
The typical rms noise level is 0.05\mjb. \label{fig:super}}
\end{figure*}

\begin{figure*}[!ht]
\begin{center}
\includegraphics[width=0.35\textwidth,bb=162 200 600 566,clip=]{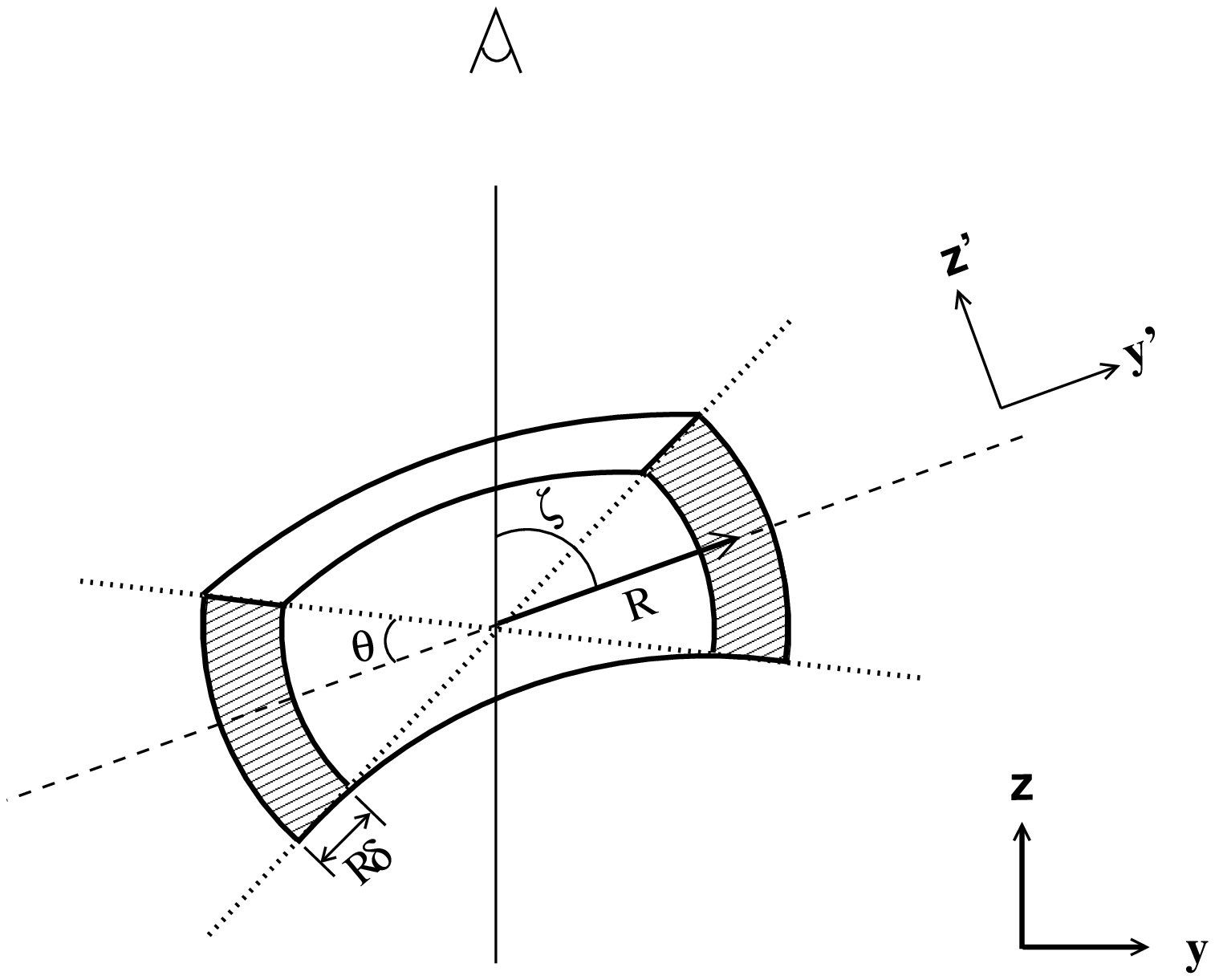}
\hspace*{5mm}
\includegraphics[width=0.31\textwidth,clip=]{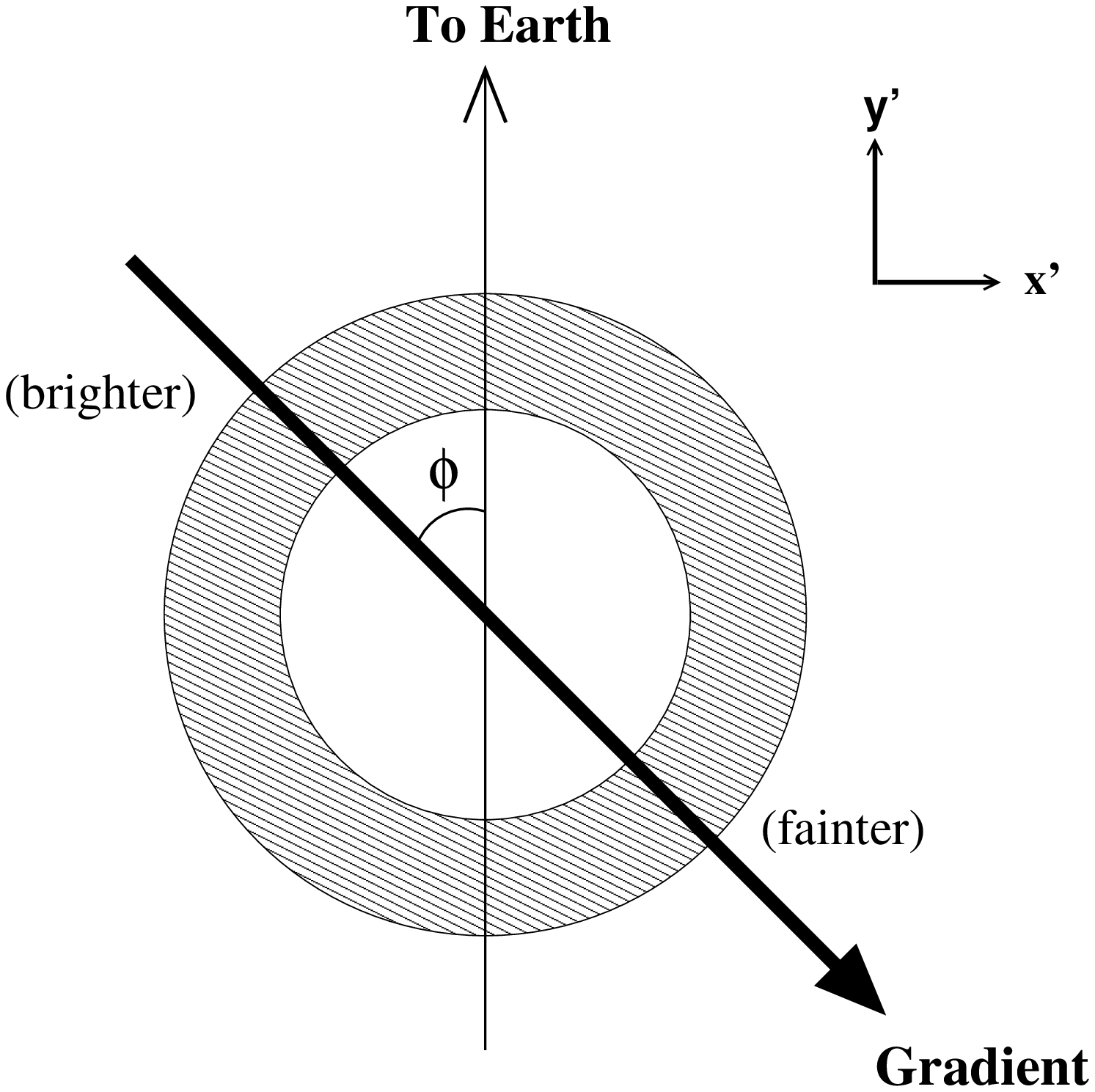}
\hspace*{5mm}
\includegraphics[width=0.23\textwidth,bb= 0 -30 319 234]{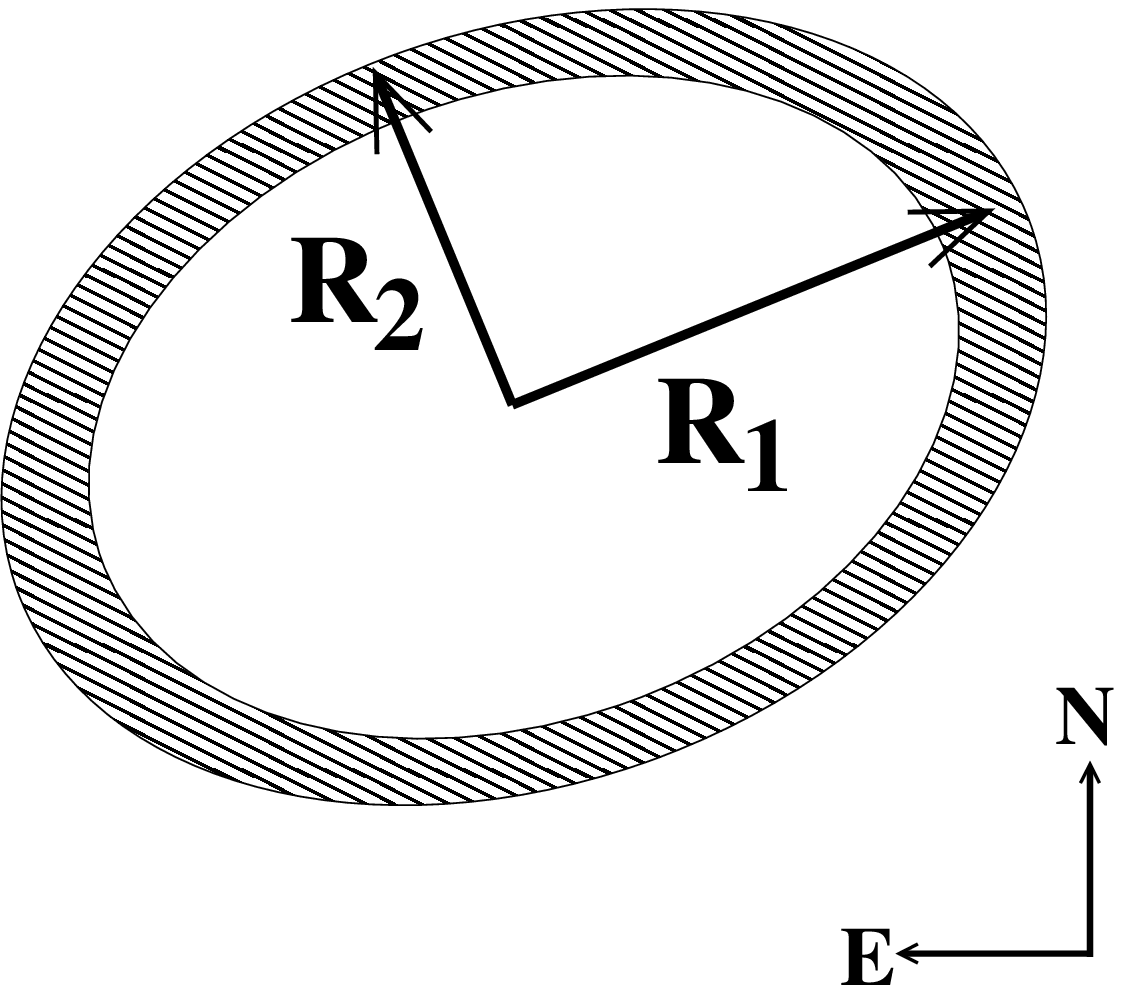}
\end{center}
\vspace*{-5mm}
\caption{Illustration of the torus and ring models used in Fourier modeling.
The left and middle panels are adopted from \citetalias{ngs+08} to show the
torus parameters: radius ($R$), half-opening angle ($\theta$), inclination
angle ($\zeta$), thickness ($R\delta$), position angle ($\phi$) of the linear
gradient in emissivity. The right panel shows the elliptical ring model in the
image plane, with semi-major and semi-minor axes of $R_1$ and $R_2$,
respectively. The ring's thickness is assumed to be negligible and the
position angle is fixed during the fit. \label{fig:model}}
\vspace*{-5mm}
\end{figure*}

ATCA imaging observations at 9\,GHz have been taken about twice a year since
1992 \citep[hereafter \citetalias{ngs+08}]{gms+97,mgw+02,ngs+08}. With the
source flux increase and various upgrades to the telescope, the remnant has
been resolved at progressively higher frequencies, from 18\,GHz to 36\,GHz to
44\,GHz to 94\,GHz \citep{mgs+05,psn+09,zsn+13,lzv+12}, and also with very
long baseline interferometry VLBI at 1.4\,GHz and 1.7\,GHz
\citep{tpa+09,nps+11}. The remnant shows similar structure at all these
frequencies, and can be described as a thin shell with an asymmetric surface
brightness distribution along the east-west direction. A thin-shell model was
used to quantify the remnant structure in early studies \citep{smk+93,gms+97}.
\citetalias{ngs+08} developed a three-dimensional (3D) torus model that can
capture the latitude extent of the emission and the east-west asymmetry.
Fitting the torus model to observations between 1992 and 2008, a linear
expansion of $\sim4000$\kms\ was found up to day 8000, which is in contrast to
the deceleration of the X-ray remnant observed around day 6000
\citep{rpz+09,hbd+13}.

In this paper, we report on the latest evolution of the radio morphology of
SNR 1987A up to day 9568 after the SN explosion, using 9\,GHz ATCA
imaging observations taken from 1992 January to 2013 May. The observations and
Fourier modeling scheme are described in Sections~\ref{sec:data}
and~\ref{sec:model}, respectively, the modeling results are presented in
Section~\ref{sec:result} and we infer the remnant expansion rate in
Section~\ref{sec:expand}. The physical implications of the results are
discussed in Section~\ref{sec:discuss}.

\section{Observations and Data Reduction}
\label{sec:data}
Radio imaging of SNR 1987A at 9\,GHz have been carried out for 21\,yr since
1992 using ATCA in the 6-km array configuration, with a typical on-source
time of $\sim$10\,hr for each observation. In this paper, we analyze the
remnant evolution in light of datasets recorded to date, which include recent
observations taken since the Compact Array Broadband Backend
\citep[CABB;][]{wfa+11} upgrade in mid-2009. Table~\ref{table:obs} lists the
observation parameters. Pre-CABB observations were made in two frequencies
with a usable bandwidth of 104\,MHz each. While the bandwidth has been greatly
increased to 2\,GHz since the CABB upgrade, we restricted our analysis of the
CABB data in two sub-bands only, with center frequencies of 8.512\,GHz and
8.896\,GHz and a bandwidth of 104\,MHz each, for a consistent comparison with
previous data. Note that the dataset taken on 2012 December 9 was affected by a
storm, leaving only 6\,hr of on-source time. It was not used in this study,
because the $u$-$v$ sampling is inadequate for high-fidelity imaging,
resulting in large rms noise in the final maps (0.5\mjb)

All data were reduced using the MIRIAD package \citep{stw95}. After standard
flagging and calibration, we employed self-calibration on data taken after
1996, when the source was detected with a high signal-to-noise ratio (see
\citealp{gms+97}; \citetalias{ngs+08}). The visibility data were averaged with
five-minute intervals. The intensity maps formed from the visibility were
deconvolved using a maximum entropy algorithm \citep{gd78}. We applied a
super-resolution technique \citep{sbr+93} by restoring the cleaned maps with a
super-resolved circular beam of FWHM 0\farcs4 (see \citetalias{ngs+08} for
the details of the reduction procedure). The final maps have a typical rms
noise of 0.05\mjb. Figure~\ref{fig:super} shows the resulting images. The
remnant morphology can be described by a circular shell with two bright lobes.
The eastern lobe is always brighter than the western lobe. Over the epochs,
the remnant has brightened significantly and exhibited a clear expansion. 

\begin{deluxetable*}{cccccccc}
\tablecaption{Best-fit Parameters for the Torus Model with Statistical
Uncertainties at 68\% Confidence Level\label{table:torus}}
\tablewidth{0pt}
\tabletypesize{\small}
\tablehead{
\colhead{Day}&\colhead{Flux (mJy)}&\colhead{Radius (\arcsec)}
&\colhead{Half-opening}&\colhead{Thickness}&\colhead{Asymmetry}&\colhead{$\phi$ (\arcdeg)}&\colhead{$\chi^2_\nu$/dof\tablenotemark{a}}\\
&&&Angle (\arcdeg)&\colhead{(\%)}&\colhead{(\%)}}
\startdata
1786& $4.2\pm0.2$ & $0.60\pm0.10$ & $84^{+6}_{-20} $ & $150\pm50$ & $70\pm30$ & $187\pm16$ &1.8/2107\\
1852& $4.0\pm0.3$ & $0.62\pm0.05$ & $80\pm10$ & $100\pm50$ & $100_{-60} $ & $180^{+10}_{-30}$ &3.7/1642\\
2067& $5.73\pm0.12$ & $0.62\pm0.05$ & $33\pm4$ & $175\pm25$ & $81\pm7$ & $121\pm6$ &4.3/3602\\
2142& $5.32\pm0.14$ & $0.65\pm0.02$ & $44\pm2$ & $172\pm10$ & $96\pm3$ & $114\pm7$ &17/2702\\
2143& $5.7\pm0.2$ & $0.64\pm0.01$ & $0^{+12} $ &$0^{+20} $ & $40\pm8$ & $108\pm6$ &16/1392\\
2313& $6.73\pm0.11$ & $0.63\pm0.01$ & $34\pm7$ & $0^{+20} $ & $40\pm5$ & $94\pm14$ &3.6/2902\\
2320& $7.04\pm0.13$ & $0.67\pm0.02$ & $37\pm7$ & $44^{+15}_{-20} $ & $38\pm5$ & $95\pm13$ &4.5/2962\\
2426& $6.65\pm0.10$ & $0.69\pm0.01$ & $55\pm4$ & $33^{+10}_{-16} $ & $42\pm5$ & $85\pm12$ &5.2/4372\\
2550& $6.63\pm0.15$ & $0.64\pm0.04$ & $26\pm5$ & $175\pm20$ & $80\pm6$ & $108\pm5$ &7.5/2992\\
2681& $8.41\pm0.08$ & $0.67\pm0.01$ & $48\pm3$ & $18^{+10}_{-17} $ & $33\pm4$ & $92\pm10$ &6.0/6142 \\
2685& $8.11\pm0.10$ & $0.66\pm0.01$ & $54\pm4$ & $0^{+14} $ & $38\pm6$ & $113\pm12$ &5.7/3256\\
3073& $11.11\pm0.12$ & $0.67\pm0.01$ & $34\pm5$ & $46\pm14$ & $40\pm3$ & $93\pm8$ &5.6/2662\\
3109& $9.7\pm0.1$ & $0.64\pm0.02$ & $18^{+10}_{-18} $ & $0^{+15} $ & $42\pm2$ & $88\pm7$ &17/1598 \\
3178& $11.71\pm0.09$ & $0.685\pm0.007$ & $45\pm2$ & $0^{+10} $ & $39\pm2$ & $103\pm7$ &4/3442\\
3436& $15.17\pm0.09$ & $0.705\pm0.005$ & $47\pm2$ & $24\pm11$ & $42\pm2$ & $95\pm4$ &4.9/4337\\
3485& $15.42\pm0.08$ & $0.707\pm0.005$ & $51\pm2$ & $1^{+10} $ & $43\pm2$ & $102\pm5$ &5.2/4702\\
3512& $15.43\pm0.12$ & $0.708\pm0.006$ & $53\pm2$ & $0^{+18} $ & $42\pm3$ & $94\pm6$ &3.3/2632\\
3914& $17.57\pm0.14$ & $0.694\pm0.007$ & $42\pm3$ & $0^{+10} $ & $38\pm3$ & $111\pm7$ &2.4/1272\\
4013& $19.09\pm0.10$ & $0.754\pm0.005$ & $46.4\pm1.4$ & $0^{+5} $ & $45.1\pm1.5$ & $104\pm4$ &3.0/2830\\
4016& $18.72\pm0.10$ & $0.745\pm0.006$ & $51\pm2$ & $0^{+5} $ & $46\pm2$ & $103\pm5$ &3.1/2512\\
4220& $20.20\pm0.09$ & $0.729\pm0.004$ & $43.4\pm1.3$ & $2^{+13}_{-2} $ & $37.6\pm1.3$ & $100\pm3$ &2.2/2955\\
4268& $21.78\pm0.13$ & $0.736\pm0.006$ & $40\pm2$ & $28\pm8$ & $38\pm2$ & $107\pm4$ &7.3/3862\\
4372& $22.94\pm0.10$ & $0.727\pm0.005$ & $37\pm2$ & $23\pm9$ & $37.4\pm1.5$ & $114\pm3$ &3.6/3532
\\
4577& $23.89\pm0.14$ & $0.757\pm0.007$ & $40\pm2$ & $21\pm6$ & $39\pm2$ & $103\pm4$ &7.0/3442\\
4584& $25.23\pm0.07$ & $0.747\pm0.003$ & $42.0\pm1.0$ & $0^{+5} $ & $38.5\pm1.0$ & $109\pm2$ &2.9/4222\\
4966& $29.45\pm0.06$ & $0.764\pm0.002$ & $40.8\pm0.6$ & $0^{+5} $ & $39.5\pm0.6$ & $108.3\pm1.3$ &1.3/50689\\
5011& $32.97\pm0.07$ & $0.775\pm0.002$ & $44.1\pm0.7$ & $1^{+5}_{-1} $ & $40.1\pm0.6$ & $104.9\pm1.4$ &1.4/50531\\
5387& $34.11\pm0.08$ & $0.790\pm0.003$ & $41.4\pm0.8$ & $0^{+3} $ & $41.5\pm0.7$ & $107.8\pm1.5$ &1.6/39604\\
5748& $41.68\pm0.07$ & $0.811\pm0.002$ & $43.8\pm0.5$ & $0^{+2} $ & $40.2\pm0.5$ & $103.1\pm1.2$ &1.3/38992\\
5810& $42.46\pm0.07$ & $0.815\pm0.002$ & $42.9\pm0.5$ & $1^{+4}_{-1} $ & $42.8\pm0.6$ & $117.4\pm1.0$ &1.6/46012\\
6003& $46.50\pm0.07$ & $0.815\pm0.002$ & $39.6\pm0.5$ & $0^{+2} $ & $38.4\pm0.5$ & $101.2\pm1.1$ &1.6/44992\\
6129& $52.82\pm0.08$ & $0.833\pm0.002$ & $42.7\pm0.5$ & $0^{+2} $ & $42.1\pm0.4$ & $107.7\pm1.1$ &1.3/46012\\
6170& $54.05\pm0.08$ & $0.831\pm0.002$ & $42.2\pm0.5$ & $0^{+2} $ & $38.8\pm0.4$ & $107.8\pm1.1$ &1.5/43672\\
6283& $53.63\pm0.07$ & $0.829\pm0.001$ & $39.6\pm0.4$ & $0^{+2} $ & $38.8\pm0.4$ & $107.7\pm0.9$ &1.5/44842\\
6605& $61.36\pm0.10$ & $0.843\pm0.002$ & $38.2\pm0.5$ & $0^{+2} $ & $39.1\pm0.5$ & $109.0\pm1.1$ &2.9/42892\\
6693& $62.69\pm0.08$ & $0.858\pm0.001$ & $43.0\pm0.4$ & $0^{+3} $ & $35.9\pm0.4$ & $101.5\pm1.0$ &1.5/38992\\
6973& $73.81\pm0.08$ & $0.880\pm0.001$ & $44.6\pm0.3$ & $0^{+2} $ & $42.1\pm0.4$ & $117.4\pm0.7$ &1.5/40357\\
7085& $77.19\pm0.08$ & $0.872\pm0.001$ & $39.3\pm0.3$ & $0^{+1} $ & $39.8\pm0.3$ & $111.8\pm0.6$ &1.5/29112\\
7228& $82.51\pm0.08$ & $0.874\pm0.001$ & $40.0\pm0.3$ & $0^{+1} $ & $39.4\pm0.3$ & $109.1\pm0.7$ &1.4/35677\\
7620& $93.61\pm0.09$ & $0.893\pm0.001$ & $42.7\pm0.3$ & $0^{+4} $ & $38.9\pm0.3$ & $105.1\pm0.6$ &1.3/42892\\
7730& $98.98\pm0.08$ & $0.8905\pm0.0008$ & $36.2\pm0.2$ & $0^{+1} $ & $36.3\pm0.2$ & $109.4\pm0.5$ &1.6/40942\\
7901& $107.73\pm0.08$ & $0.8916\pm0.0007$ & $35.8\pm0.2$ & $0^{+2} $ & $35.9\pm0.2$ & $109.1\pm0.4$ &1.5/44452\\
8139& $121.59\pm0.08$ & $0.9095\pm0.0006$ & $37.6\pm0.1$ & $0^{+4} $ & $32.0\pm0.1$ & $104.4\pm0.4$ &0.5/278820\\
8370& $128.07\pm0.08$ & $0.9142\pm0.0006$ & $36.6\pm0.2$ & $0^{+1} $ & $33.4\pm0.2$ & $105.4\pm0.4$ &0.8/254302\\
8448& $132.59\pm0.07$ & $0.9109\pm0.0008$ & $33.5\pm0.2$ & $12\pm3$ & $28.7\pm0.1$ & $103.4\pm0.4$ &0.9/301177\\
8737& $142.25\pm0.07$ & $0.9185\pm0.0005$ & $32.8\pm0.1$ & $0^{+2} $ & $29.8\pm0.1$ & $104.2\pm0.4$ &0.6/251707\\
8824& $136.58\pm0.08$ & $0.9169\pm0.0008$ & $28.4\pm0.2$ & $1^{+3}_{-1} $ & $26.9\pm0.1$ & $105.7\pm0.4$ &0.7/241327\\
9089& $155.92\pm0.06$ & $0.9251\pm0.0005$ & $30.5\pm0.1$ & $0^{+1} $ & $25.8\pm0.1$ & $102.3\pm0.4$ &1.1/265987\\
9233& $161.00\pm0.06$ & $0.9290\pm0.0006$ & $28.8\pm0.1$ & $1^{+2}_{-1} $ & $25.8\pm0.1$ & $102.4\pm0.3$ &1.0/301177\\
9321& $165.62\pm0.05$ & $0.9304\pm0.0003$ & $29.1\pm0.1$ & $0^{+1} $ & $23.5\pm0.1$ & $98.4\pm0.3$ &0.9/284272\\
9509& $168.68\pm0.05$ & $0.9467\pm0.0003$ & $31.7\pm0.1$ & $0^{+1} $ & $26.7\pm0.1$ & $97.8\pm0.2$ &1.4/288550\\
9568& $176.01\pm0.04$ & $0.9378\pm0.0002$ & $28.6\pm0.1$ & $0^{+1} $ & $18.4\pm0.1$ & $111.6\pm0.3$ &1.3/283341
\enddata
\tablenotetext{a}{Before 2000, all 26 frequency channels in the data were
averaged into one band of effective bandwidth 208\,MHz to boost the signal;
between 2000 and 2009, 26 Hanning-smoothed channels, each of width 8\,MHz were
used in the fit; since mid-2009, after the installation of the Compact Array
Broadband Backend (CABB), 208 channels in the same frequency range were extracted,
each of width 1\,MHz. Since 2012, the ATCA sensitivity has improved by
$\sim$40\% as a result of the installation of new receivers.}
\end{deluxetable*}

\begin{deluxetable*}{cccccccc}
\tablecaption{Best-fit Parameters for the Ring Model with Statistical
Uncertainties at 68\% Confidence Level\label{table:ring}}
\tablewidth{0pt}
\tabletypesize{\small}
\tablehead{
\colhead{Day}&\colhead{Flux (mJy)}&\colhead{Semi-major}&
\colhead{Semi-minor}&\colhead{Asymmetry}&\colhead{$\phi$
(\arcdeg)}&\colhead{$\chi^2_\nu$/dof\tablenotemark{a}}\\
&&\colhead{Axis (\arcsec)}&\colhead{Axis (\arcsec)}&
\colhead{(\%)} &}
\startdata
1786& $3.70\pm0.12$ & $0.55\pm0.03$ & $0.50\pm0.03$ & $33\pm16$ & $141\pm26$ &1.8/2108\\
1852& $3.59\pm0.14$ & $0.53\pm0.04$ & $0.48\pm0.04$ & $25\pm17$ & $124\pm46$ &3.7/1643\\
2067& $5.17\pm0.09$ & $0.57\pm0.02$ & $0.49\pm0.01$ & $27\pm5$ & $105\pm16$ &4.4/3603\\
2142& $4.84\pm0.11$ & $0.53\pm0.02$ & $0.51\pm0.02$ & $17\pm8$ & $132\pm29$ &17/2703\\
2143& $5.62\pm0.14$ & $0.63\pm0.03$ & $0.44\pm0.02$ & $39\pm8$ & $94\pm24$ &16/1393\\
2313& $6.68\pm0.09$ & $0.57\pm0.02$ & $0.44\pm0.01$ & $34\pm5$ & $90\pm15$ &3.6/2903\\
2320& $6.87\pm0.10$ & $0.61\pm0.02$ & $0.47\pm0.01$ & $31\pm4$ & $92\pm13$ &4.5/2963\\
2426& $6.47\pm0.08$ & $0.56\pm0.01$ & $0.50\pm0.01$ & $28\pm4$ & $86\pm9$ &5.2/4373\\
2550& $6.12\pm0.12$ & $0.58\pm0.02$ & $0.54\pm0.02$ & $32\pm6$ & $117\pm12$ &7.5/2993\\
2681& $8.23\pm0.06$ & $0.58\pm0.01$ & $0.48\pm0.01$ & $26\pm3$ & $87\pm8$ &6.0/6143\\
2685& $7.94\pm0.08$ & $0.56\pm0.01$ & $0.49\pm0.01$ & $29\pm4$ & $109\pm10$ &5.7/3257\\
3073& $10.86\pm0.10$ & $0.62\pm0.01$ & $0.47\pm0.01$ & $33\pm2$ & $87\pm8$ &5.6/2663\\
3109& $9.73\pm0.23$ & $0.60\pm0.01$ & $0.50\pm0.01$ & $36\pm8$ & $92\pm32$ &17/1599\\
3178& $11.50\pm0.07$ & $0.601\pm0.007$ & $0.487\pm0.006$ & $32\pm2$ & $98\pm6$ &4.0/3443\\
3436& $14.76\pm0.07$ & $0.607\pm0.005$ & $0.495\pm0.004$ & $33\pm1$ & $94\pm4$ &5.0/4338\\
3485& $15.07\pm0.07$ & $0.598\pm0.006$ & $0.510\pm0.004$ & $33\pm2$ & $98\pm4$ &5.2/4703\\
3512& $15.06\pm0.10$ & $0.592\pm0.007$ & $0.508\pm0.005$ & $32\pm2$ & $91\pm5$ &3.4/2633\\
3914& $17.31\pm0.12$ & $0.624\pm0.006$ & $0.491\pm0.007$ & $33\pm3$ & $107\pm7$ &2.4/1273\\
4013& $18.53\pm0.09$ & $0.646\pm0.005$ & $0.527\pm0.004$ & $35.5\pm1.2$ & $99\pm3$ &3.0/2831\\
4016& $18.36\pm0.09$ & $0.632\pm0.005$ & $0.545\pm0.006$ & $35.8\pm1.3$ & $98\pm4$ &3.1/2513\\
4220& $19.76\pm0.07$ & $0.643\pm0.004$ & $0.509\pm0.003$ & $31.4\pm1.0$ & $96\pm3$ &2.2/2956\\
4268& $21.19\pm0.11$ & $0.656\pm0.006$ & $0.506\pm0.005$ & $31.1\pm1.5$ & $104\pm4$ &7.4/3863\\
4372& $22.43\pm0.08$ & $0.663\pm0.004$ & $0.500\pm0.004$ & $32.1\pm1.2$ & $111\pm3$ &3.6/3533\\
4577& $23.26\pm0.12$ & $0.674\pm0.006$ & $0.515\pm0.005$ & $32.7\pm1.4$ & $98\pm4$ &7.1/3443\\
4584& $24.69\pm0.07$ & $0.663\pm0.003$ & $0.521\pm0.003$ & $31.6\pm0.8$ & $103\pm2$ &2.9/4223\\
4966& $28.75\pm0.05$ & $0.681\pm0.002$ & $0.527\pm0.002$ & $32.9\pm0.5$ & $102\pm1$ &1.3/50690\\
5011& $32.03\pm0.06$ & $0.677\pm0.002$ & $0.531\pm0.002$ & $32.7\pm0.5$ & $100\pm1$ &1.4/50532\\
5387& $33.19\pm0.07$ & $0.694\pm0.002$ & $0.547\pm0.002$ & $34.0\pm0.5$ & $102\pm1$ &1.6/39605\\
5748& $40.56\pm0.06$ & $0.710\pm0.002$ & $0.564\pm0.001$ & $33.0\pm0.4$ & $97\pm1$ &1.3/38993\\
5810& $41.24\pm0.06$ & $0.723\pm0.002$ & $0.568\pm0.002$ & $34.8\pm0.4$ & $110\pm1$ &1.6/46013\\
6003& $45.35\pm0.06$ & $0.723\pm0.002$ & $0.561\pm0.002$ & $32.2\pm0.4$ & $94\pm1$ &1.6/44993\\
6129& $51.43\pm0.07$ & $0.733\pm0.002$ & $0.579\pm0.002$ & $34.7\pm0.3$ & $100\pm1$ &1.3/46013\\
6170& $52.63\pm0.07$ & $0.734\pm0.001$ & $0.577\pm0.002$ & $32.1\pm0.3$ & $102\pm1$ &1.5/43673\\
6283& $52.18\pm0.06$ & $0.743\pm0.001$ & $0.569\pm0.001$ & $32.8\pm0.3$ & $101\pm1$ &1.5/44843\\
6605& $59.86\pm0.09$ & $0.758\pm0.002$ & $0.586\pm0.002$ & $33.0\pm0.4$ & $101\pm1$ &2.9/42893\\
6693& $61.21\pm0.07$ & $0.7622\pm0.0013$ & $0.5979\pm0.0013$ & $30.1\pm0.3$ & $95\pm1$ &1.5/38993\\
6973& $71.33\pm0.07$ & $0.7743\pm0.0011$ & $0.6118\pm0.0014$ & $33.7\pm0.3$ & $110\pm1$ &1.5/40358\\
7085& $74.60\pm0.07$ & $0.7695\pm0.0012$ & $0.5939\pm0.0011$ & $32.6\pm0.2$ & $103\pm1$ &1.6/29113\\
7228& $80.53\pm0.07$ & $0.7825\pm0.0009$ & $0.6171\pm0.0010$ & $33.0\pm0.2$ & $101\pm1$ &1.4/35678\\
7620& $90.47\pm0.08$ & $0.7830\pm0.0011$ & $0.6208\pm0.0009$ & $31.8\pm0.2$ & $97\pm1$ &1.4/42893\\
7730& $96.36\pm0.07$ & $0.8058\pm0.0008$ & $0.6179\pm0.0008$ & $30.9\pm0.2$ & $102\pm1$ &1.6/40943\\
7901& $104.94\pm0.06$ & $0.8075\pm0.0007$ & $0.6220\pm0.0007$ & $30.8\pm0.1$ & $102\pm0$ &1.5/44453\\
8139& $117.73\pm0.07$ & $0.8241\pm0.0006$ & $0.6263\pm0.0006$ & $27.7\pm0.1$ & $97\pm0$ &0.5/278821 \\
8370& $124.86\pm0.07$ & $0.8243\pm0.0006$ & $0.6428\pm0.0007$ & $28.6\pm0.1$ & $98\pm0$ &0.8/254303 \\
8448& $129.50\pm0.06$ & $0.8406\pm0.0005$ & $0.6282\pm0.0006$ & $25.6\pm0.1$ & $94\pm0$ &0.9/301178 \\
8737& $139.17\pm0.06$ & $0.8433\pm0.0005$ & $0.6430\pm0.0005$ & $26.2\pm0.1$ & $96\pm0$ &0.7/251708 \\
8824& $133.95\pm0.07$ & $0.8582\pm0.0006$ & $0.6373\pm0.0007$ & $24.1\pm0.1$ & $96\pm0$ &0.7/241328 \\
9089& $153.09\pm0.05$ & $0.8593\pm0.0004$ & $0.6438\pm0.0005$ & $23.1\pm0.1$ & $92\pm0$ &1.1/265988 \\
9233& $157.94\pm0.05$ & $0.8698\pm0.0004$ & $0.6401\pm0.0005$ & $23.4\pm0.1$ & $93\pm0$ &1.1/301178 \\
9321& $162.98\pm0.05$ & $0.8703\pm0.0003$ & $0.6527\pm0.0004$ & $21.5\pm0.1$ & $88\pm0$ &0.9/284273 \\
9509& $165.10\pm0.04$ & $0.8761\pm0.0003$ & $0.6583\pm0.0003$ & $24.0\pm0.1$ & $88\pm0$ &1.4/288551 \\
9568& $173.03\pm0.04$ & $0.8818\pm0.0002$ & $0.6574\pm0.0003$ & $16.2\pm0.1$ & $100\pm0$ &1.4/283342
\enddata 
\tablenotetext{a}{Before 2000, all 26 frequency channels in the data were
averaged into one band of effective bandwidth 208\,MHz to boost the signal;
between 2000 and 2009, 26 Hanning-smoothed channels, each of width 8\,MHz were
used in the fit; since mid-2009, after the installation of the Compact Array
Broadband Backend (CABB), 208 channels in the same frequency range were extracted,
each of width 1\,MHz. Since 2012, the ATCA sensitivity has improved by
$\sim$40\% as a result of the installation of new receivers.}
\end{deluxetable*}

\section{Fourier Modeling}
\label{sec:model}
We followed previous studies (e.g., \citealp{smk+93,gms+97};
\citetalias{ngs+08}) to assess the remnant geometry in the $u$-$v$ plane with
the Fourier modeling technique. This can give more robust measurements than
directly fitting a model in the image plane, since the Fourier domain is
where the visibility data are taken and the measurement errors are
uncorrelated. We employed two models in this study: a 3D torus as developed by
\citetalias{ngs+08} and a two-dimensional (2D) thin elliptical ring.

\begin{figure*}[!th]
\includegraphics[height=\textwidth,angle=-90]{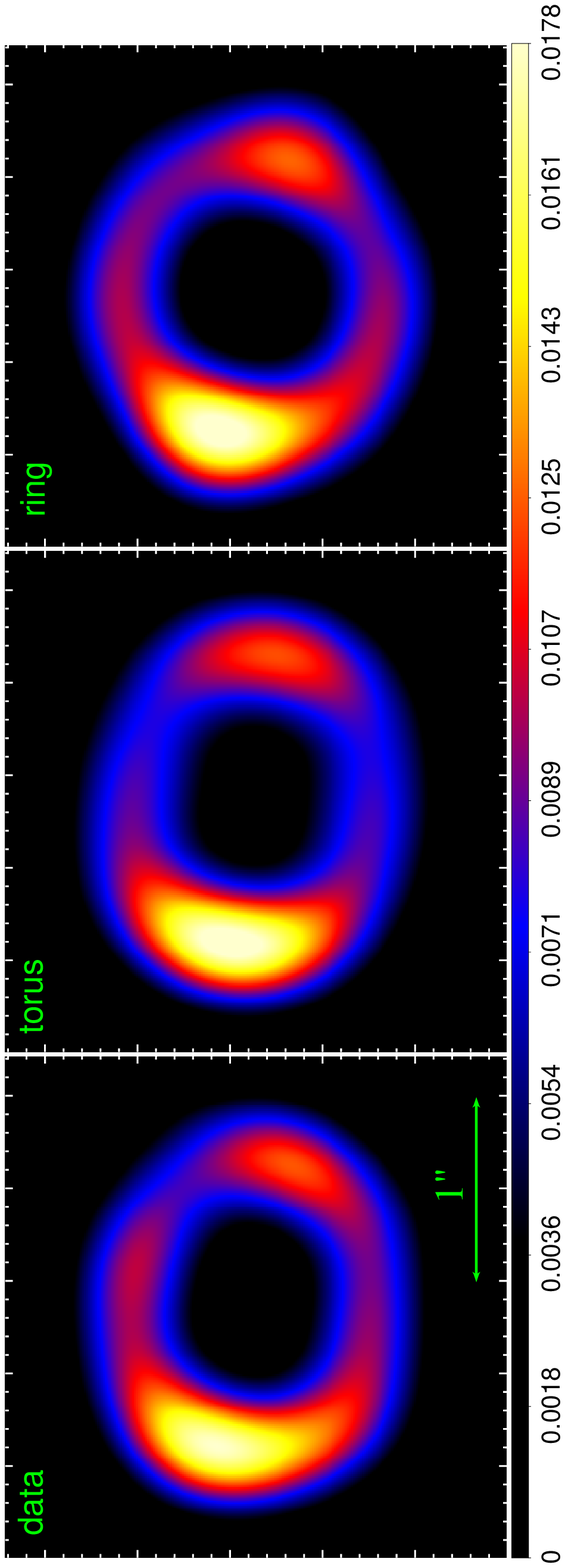}
\hspace*{0.326\textwidth}\includegraphics[height=0.667\textwidth,angle=-90]{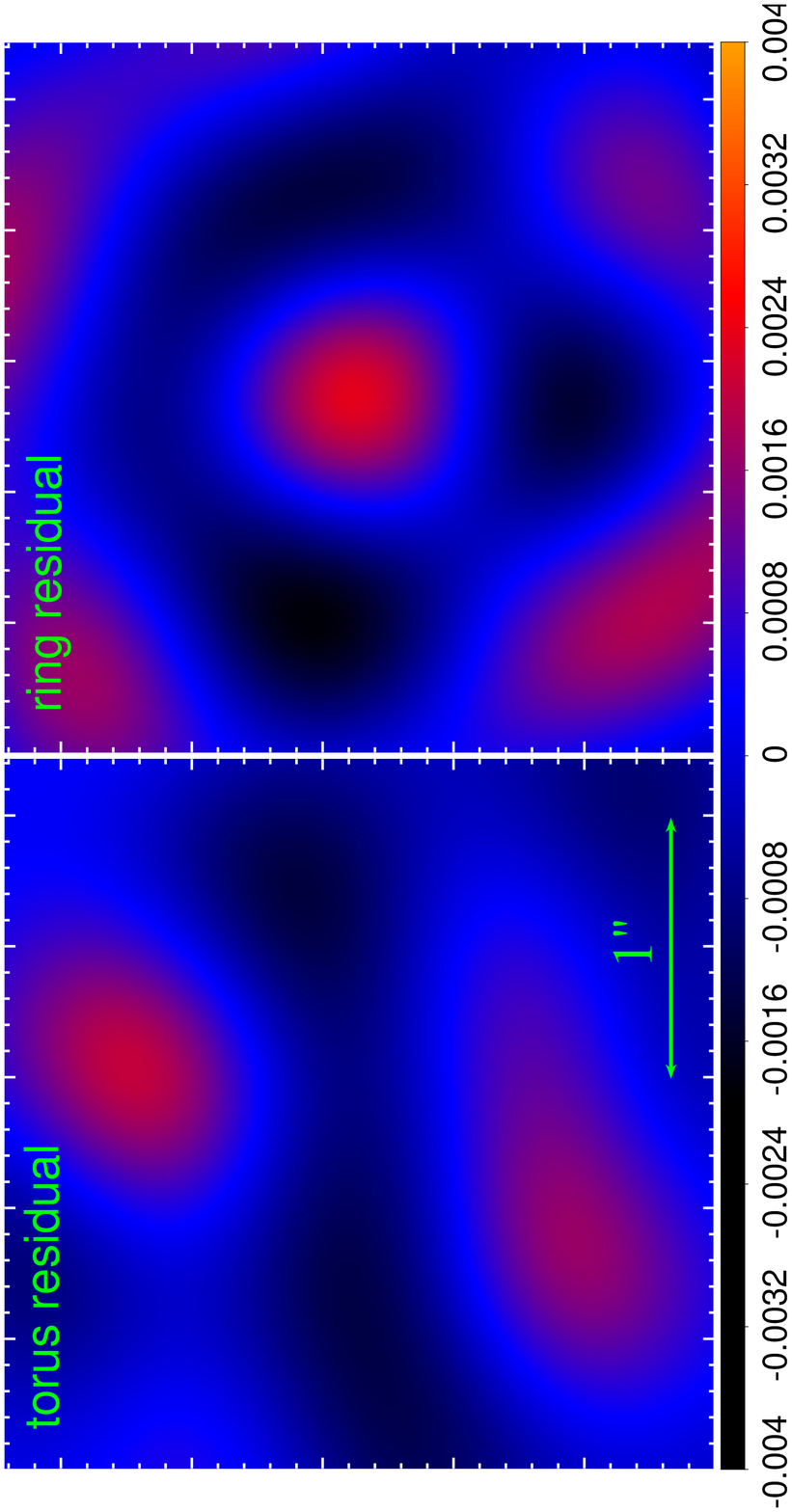}
\caption{Radio image of SN 1987A at 9\,GHz taken on 2012 September 1 (left)
compared with images of the best-fit torus (middle) and elliptical ring
(right) models. The lower panels show the maps of the residual (i.e.\
data minus model) visibilities, for which no deconvolution has been
applied. Note that the upper and lower panels have different color
scales.\hspace{\textwidth}(A color version of this figure is available in the
online journal.) \label{fig:model_img}}
\end{figure*}

The torus model is illustrated in Figure~\ref{fig:model}. It is a truncated
shell with eight fitting parameters: flux density, center position (right ascension
and declination), radius, half-opening angle ($\theta$), thickness (as a fraction
of the radius), slope (in \%) and direction ($\phi$) of a linear gradient in
the surface emissivity. The actual fitting was performed with a modified
version of the MIRIAD task \texttt{UVFIT} (see \citetalias{ngs+08} for
details). As an update to the modeling, we employed the 9\,GHz light curve
reported by \citet{zsb+10} to account for the time-of-flight effect that
causes the near side of the remnant to appear brighter than the far side. This
gives slightly different results compared to \citetalias{ngs+08}, however, we
emphasize that the differences are minimal and within the uncertainties. 

While \citetalias{ngs+08} obtained confidence intervals of the fitting
parameters using a simple bootstrapping technique, we note that the bootstrap
samples may not be representative of the true population, and other
techniques, such as subsample bootstrapping, are needed \citep[see][]{km05b}.
To avoid complications, we determined the confidence intervals from the
covariance matrices. There are a few cases for which this method fails to give
sensible results because the best-fit parameters are near the boundaries. We
then estimated the confidence intervals by plotting out the $\chi^2$
distribution.

In addition to the torus model, an elliptical ring model was also employed to
compare with that used in X-ray studies \citep{rpz+09,hbd+13}. As in the torus
model, we included a linear gradient in the surface emissivity to account for
the observed east-west asymmetry. The ring model has seven fitting parameters:
flux density, center position, semi-major axis ($R_1$), semi-minor axis
($R_2$), and degree and direction of the gradient. The model is illustrated in
Figure~\ref{fig:model}. Since the system is known have an inclination of
$\sim$45\arcdeg\ to the line of sight \citep[e.g.,][]{sck+05}, we first tried
fitting with the ring's aspect ratio fixed at $R_2/R_1=0.7$, but found that it
gives slightly worse statistics than the torus fits. We then allowed both
$R_1$ and $R_2$ to vary, which improves the results and gives comparable
$\chi^2$ values to the torus model. Same as the torus fits, confidence
intervals of the best-fit parameters were determined from the covariance
matrix.

\section{Results}
\label{sec:result}
The best-fit model parameters are listed in Tables~\ref{table:torus} and
\ref{table:ring}. As an example, we show in Figure~\ref{fig:model_img} images
of the best-fit torus and ring models compared with the actual radio map taken
on 2012 September 1, and the maps of the residual (i.e.\ data minus model)
visibilities. The models have successfully captured the characteristic
structure of the remnant. With one extra fitting parameter, the torus model
provides slightly better fits to the data than the ring model. However, we
cannot determine whether the difference in statistics is significant, since
the standard $F$-test does not apply here. As shown in
Figure~\ref{fig:model_img}, the ring model underpredicts the flux density at
the remnant center, but provides a better fit to the northern rim than the
torus model.

\begin{figure*}[!ht]
\epsscale{1.15}
\includegraphics[width=0.53\textwidth,clip=,bb=18 144 592 640]{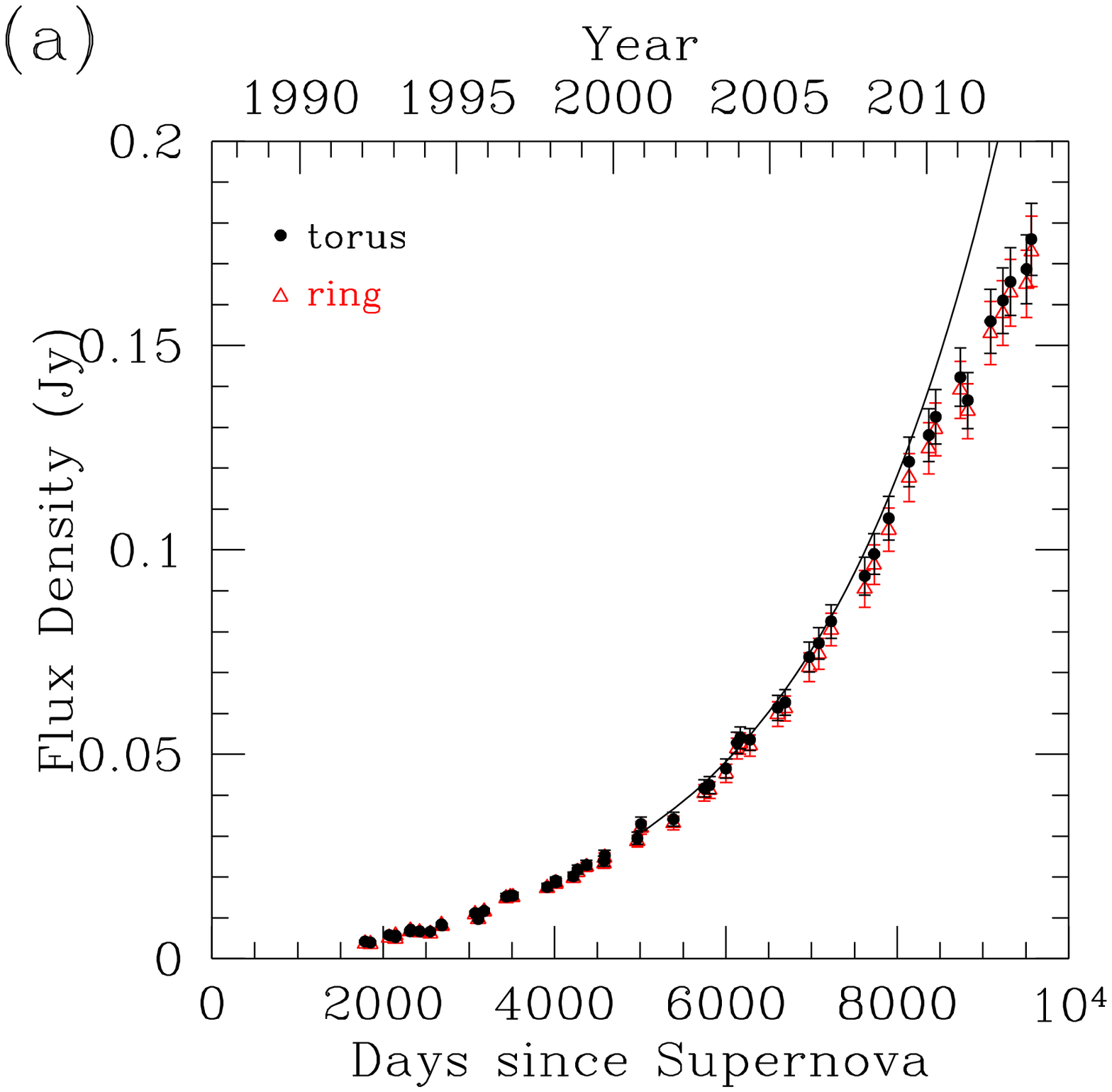}
\includegraphics[width=0.53\textwidth,clip=,bb=18 144 592 640]{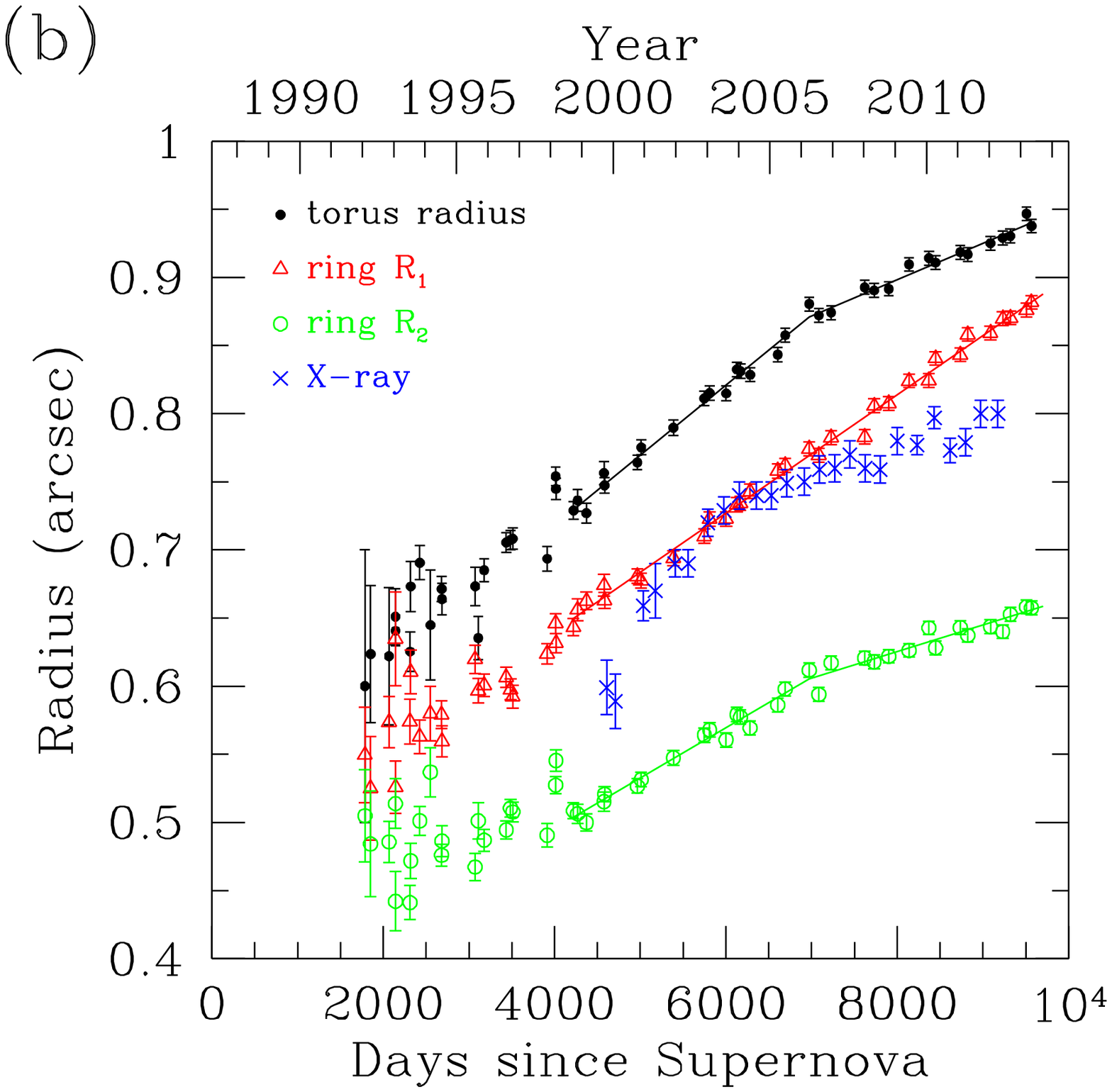}

\includegraphics[width=0.53\textwidth,clip=,bb=18 144 592 640]{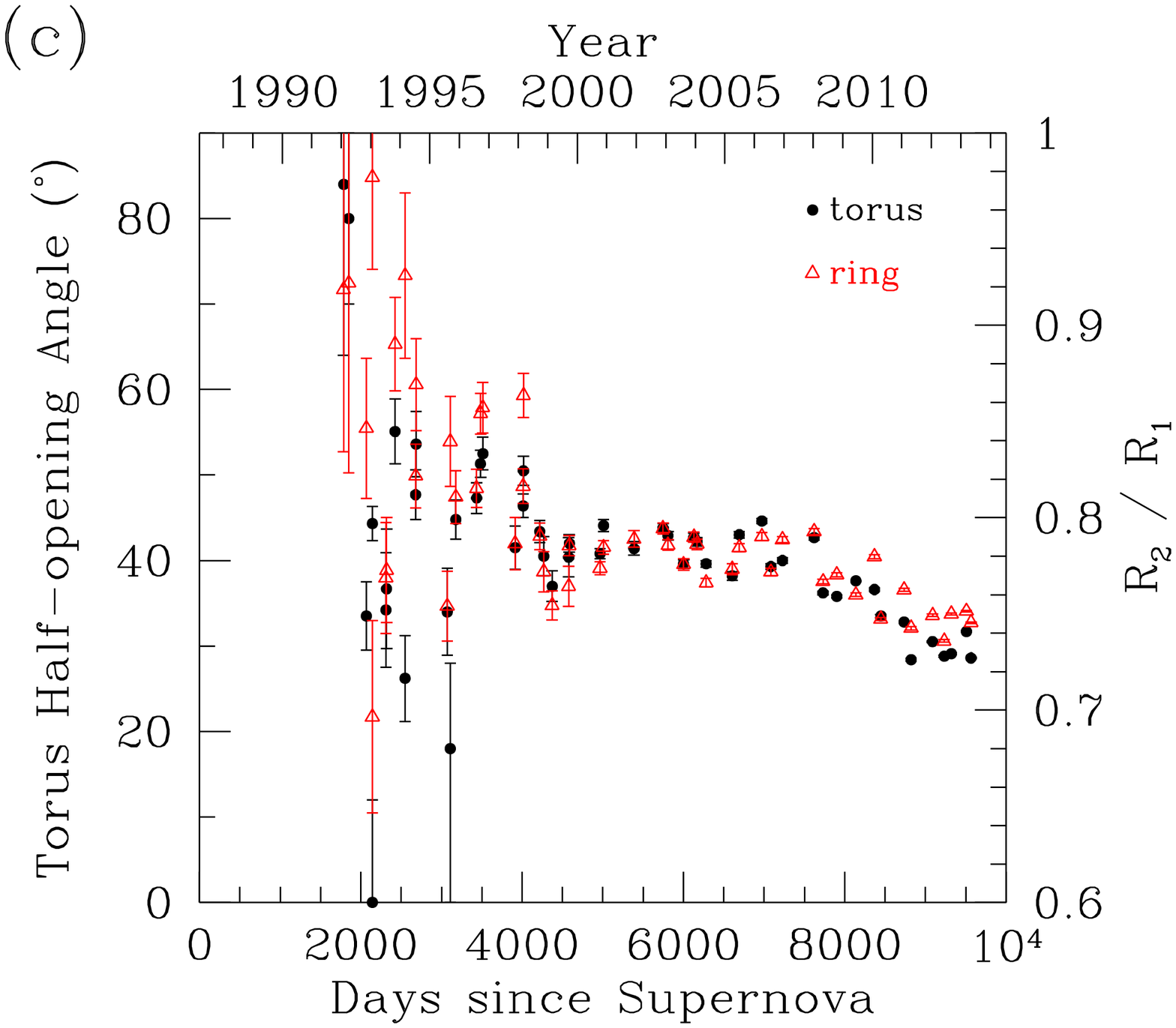}
\includegraphics[width=0.53\textwidth,clip=,bb=18 144 592 640]{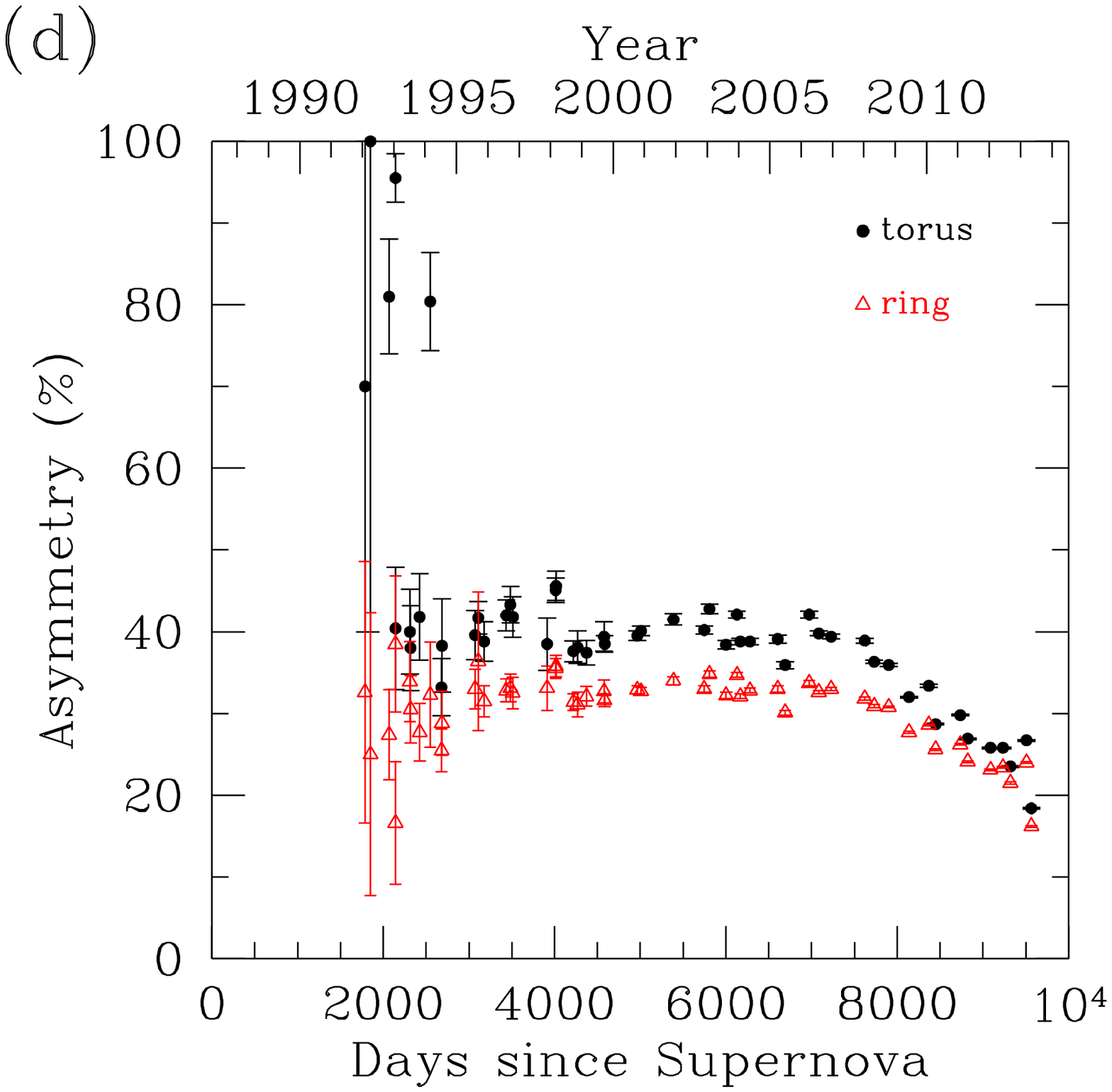}
\caption{(a) Model flux densities at 9\,GHz. Systematic uncertainties of 5\%
\citep{zsn+13} are combined with the statistical uncertainties at a 68\%
confidence level. The solid line is the exponential fit from \citet{zsb+10}.
(b) Best-fit radius of the torus model and semi-major ($R_1$) and semi-minor
($R_2$) axes of the ring model, compared with the X-ray radius reported by
\citet{hbd+13}. Systematic uncertainties of 0\farcs005 are combined with
the statistical uncertainties at 68\% confidence level. The solid lines are
the best-fit expansion rates in Table~\ref{table:expand}. (c) Best-fit torus
half-opening angle $\theta$ compared with the ratio $R_2/R_1$ of the ring
model. (d) Best-fit asymmetry in the surface brightness. \hspace{\textwidth}(A
color version of this figure is available in the online journal.) 
 \label{fig:fit}}
\end{figure*}

Time evolution of the best-fit flux density, radius, opening angle, and degree
of asymmetry, which is obtained from the slope of the gradient, are plotted in
Figure~\ref{fig:fit}. The torus radius increased linearly at the early epochs
until a clear break at day $\sim$7000. After that, the expansion rate was
significantly reduced. It is intriguing that around the same time when the break
occurred, both $\theta$ and the asymmetry began to decrease. Since day $\sim$7500
the model flux density increases at a lower rate than the exponential fit
given by \citet{zsb+10}. For the other parameters not shown in the figure, the
torus thickness is not well determined from the fits. As reported by
\citetalias{ngs+08}, this is likely smaller than the resolution of the 9\,GHz
observations \citep[see also][]{nps+11}. The direction of the linear gradient,
which identifies the asymmetry of the surface brightness, remains mostly
constant over the epochs.

\begin{figure}[th]
\epsscale{1.25}
\plotone{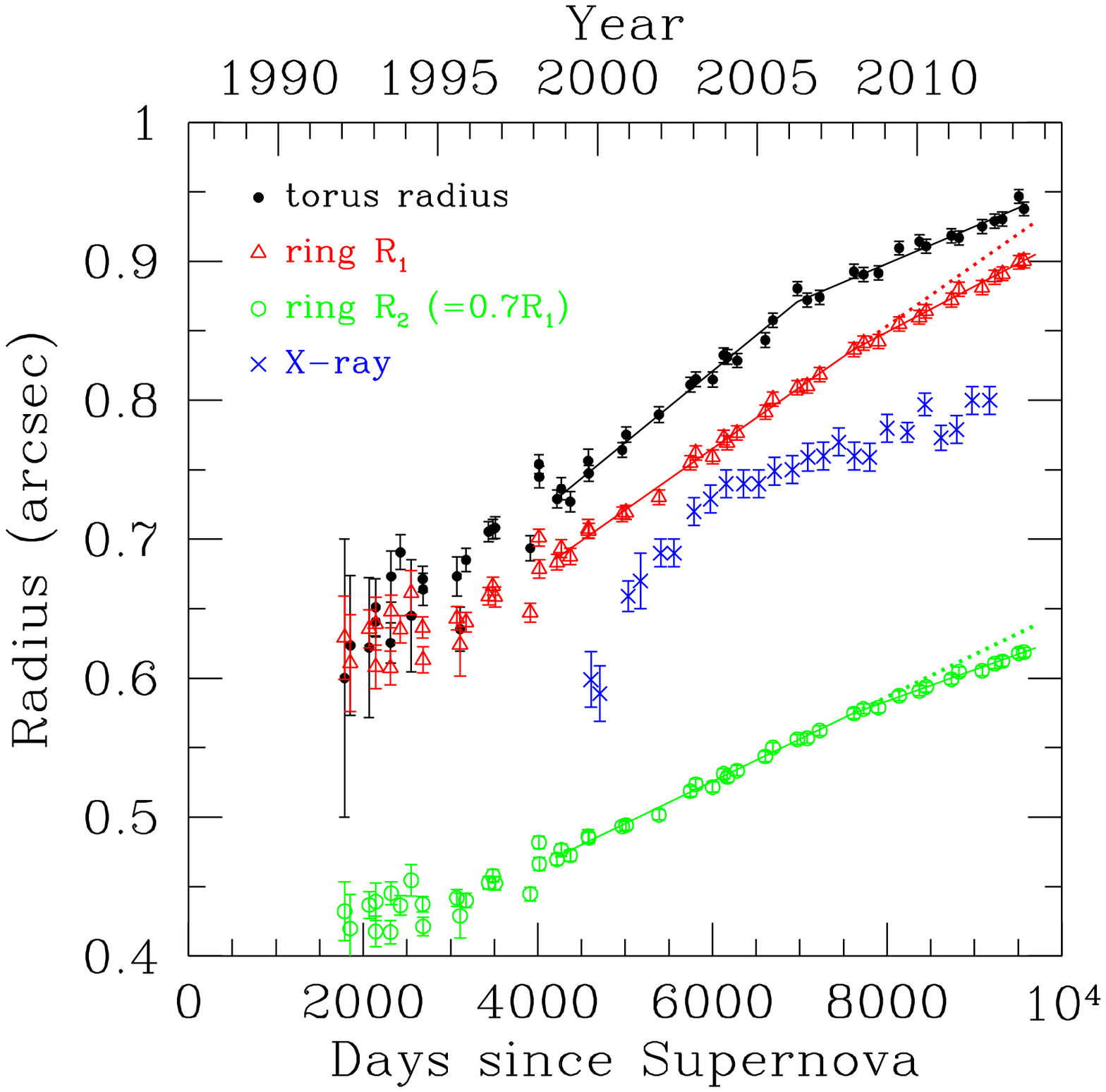}
\caption{Same Figure~(\ref{fig:fit}b), but the aspect ratio of the ring model
is fixed at $R_2/R_1$=0.7 according to the inclination of the system. To
better illustrate the break in the expansion of $R_1$ and $R_2$, the dotted
lines indicate the expansion without any breaks for a direct comparison.
\hspace{\textwidth}(A color version of this figure is available in the online
journal.) \label{fig:ringfix}}
\end{figure}

For the ring model fit, it is clear from Figure~(\ref{fig:fit}b) that $R_1$ is
always smaller than the torus radius. When compared to the X-ray radius
reported by \citet{hbd+13}, $R_1$ has a similar size between days 6000 and 7500,
and it is larger after day 7500. Over time, $R_1$ increases linearly without
obvious deceleration. In contrast, $R_2$ exhibits a break around day 7000, as
did the torus radius. This results in a decrease in the ratio $R_2/R_1$, which
shows a remarkably similar trend to that of $\theta$ (Figure~(\ref{fig:fit}c)).
The ring and torus fits give nearly identical flux densities, and both suggest
a decreasing trend in the degree of asymmetry since day $\sim$7500. For
completeness, we also show in Figure~\ref{fig:ringfix} the best-fit $R_1$ and
$R_2$ from the ring fitting with a fixed aspect ratio. We obtained larger $R_1$
and smaller $R_2$ as compared to the ring fit with varying aspect ratio, and
there is a hint of a break in the expansion around day 7500.

\begin{deluxetable*}{lccc}
\tablecaption{Expansion of the Radio Remnant of SN 1987A Since Day 4200, with
Uncertainties at 68\% Confidence Level \label{table:expand}}
\tablewidth{0pt}
\tablehead{
\colhead{Model} & \colhead{Transition Day} & \colhead{$v_1$ (\kms)} &
\colhead{$v_2$ (\kms)} }
\startdata
Torus & $7000_{-100}^{+200}$ & $4600_{-200}^{+150}$ & $2400_{-200}^{+100}$\\
Ring semi-major axis\tablenotemark{a} & \nodata & $3890\pm50$ & \nodata \\   
Ring semi-minor axis & $7000\pm300$ & $3300\pm200$ &
$1750^{+150}_{-300}$\\
Ring semi-major axis (aspect ratio fixed)\tablenotemark{b} &
$7600\pm200$ & $3940\pm70$ & $2900\pm100$\\
Ring semi-minor axis (aspect ratio fixed)\tablenotemark{b} &
$7600\pm200$ & $2710\pm50$ & $2000^{+60}_{-70}$
\enddata
\tablenotetext{a}{A simple linear expansion is preferred over a broken-linear
fit.}
\tablenotetext{b}{Expansion of the semi-major and semi-minor axes are linked
as the ring's aspect ratio is fixed during the fit.}
\end{deluxetable*} 

\section{Rate of Expansion}
\label{sec:expand}
The torus fitting results shown in Figure~(\ref{fig:fit}b) suggest a possible
deceleration of the remnant. We therefore followed \citet{rpz+09} to fit the
radius evolution with a broken-linear function. Only observations after day
4200 were used here, since we are most interested in the late evolution of the
remnant. We note that the radius measurements in Table~\ref{table:torus} have
statistical uncertainties of the order of 0\farcs001, much smaller than the
scatter from measurement to measurement. We therefore added in quadrature
systematic uncertainties of 0\farcs005 to the measurements errors. This
value was chosen to give a reduced $\chi^2$ value of about 1 for the
broken-linear fit. We have tried different values from 0 to 0\farcs05 and
confirm that our results are independent of the choice.

Table~\ref{table:expand} lists the best-fit expansion velocity and transition
day. Uncertainties quoted are 68\% confidence intervals determined using a
bootstrapping technique \citep{et93} with 10000 simulations. The break in the
expansion of the torus model occurred on day $7000^{+200}_{-100}$, and was the
same as $R_2$ of the ring model (day $7000\pm300$). However, $R_1$ shows no
breaks and a simple linear fit is statistically preferred. Finally, the ring
fit with a constant aspect ratio suggests a somewhat later transition at day
$7600\pm200$. At a source distance of 51.4\,kpc \citep{pan99}, the torus
fits suggest expansion velocities of $v_1=4600_{-200}^{+150}$\kms\ and
$v_2=2400_{-200}^{+100}$\kms\ before and after the break, respectively. While
$v_1$ is larger than the expansion rate $3890\pm50$\kms\ of $R_1$, both values
are consistent with the result of $4000\pm400$\kms\ reported by
\citetalias{ngs+08}. Similar to the torus radius, the expansion velocity of
$R_2$ decreased by nearly 50\%, from 3300\kms\ to 1750\kms, after the break.

\section{Discussion}
\label{sec:discuss} 
Non-thermal radio emission from SNRs traces the particle acceleration in
shocks. Theories suggest that the emission is generally distributed between
the forward and reverse shocks \citep[e.g.,][]{jn96}. When the shocks of SN
1987A travel outward and encounter the equatorial ring, we expect the
radio-emitting region to have a complex structure in 3D
\citep[e.g.,][]{ssn93,lms94,blc96}, with components from both high-latitude
material above the equatorial plane and the ring itself. To characterize the
remnant geometry, we have carried out Fourier modeling using a truncated-shell
torus model and an elliptical thin ring model. While both provide
adequate fits to the data, we should note that these models are simple
parameterizations of the remnant structure, and that the geometry inferred from
the fitting is model-dependent and could be systematically biased. For
instance, if the actual shape of the emission resembles a crescent torus
\citep{plc+95}, then the radius measurement obtained from our torus model
would depend sensitively on $\theta$.

One major discrepancy between the torus and ring models is the radius
measurement. We find that the radius obtained from fitting a torus model is
always larger than that from a ring model. This could be a projection effect
due to differences in the model geometry, or the emission at high latitudes,
to which the torus model is more sensitive, being physically further away
from the center. For the former, consider a thin spherical shell in 3D: in the
image the emission would appear to peak inside the shell radius because of
projection and the finite spatial resolution of the telescope. Hence, a 3D
shell model would require a larger radius compared to a simple thin ring
model. Our torus model varies between a shell and a ring depending on $\theta$
(see Figure~\ref{fig:model}). The projection effect is minimum at $\theta=0$,
where the torus reduces to a 2D ring, and increases with $\theta$. Given that
the best-fit $\theta$ is significantly greater than zero, the above
discrepancy is not unexpected. Alternatively, the radio-emitting region could
have a larger radius at higher latitudes than at the equatorial plane, because
the SN shock travels at a larger velocity in the low-density
environment \citep{blc96}. If this is the case, any 3D models sensitive to
high latitude emission, such as the torus model, will tend to give a larger
radius. It is worth noting that in both scenarios above, there is a possible
coupling between the torus radius and $\theta$, and it could correspondingly
impact the expansion measurement. 

\bibpunct[; ]{(}{)}{,}{a}{}{;} 
In previous studies, the reported radii of SNR 1987A in radio and X-rays show
a 10\% difference over day 5000--8000 (\citetalias{ngs+08}; \citealp{rpz+09}).
It was first pointed out by \citet{gsm+07} that the discrepancy may not be
physical, but due to different measuring techniques. This idea is similar to
what we have discussed above. In particular, measurements from
\citetalias{ngs+08} were made with a torus model, while the X-ray results were
obtained from ring fitting \citep{rpz+09,hbd+13}. \citet{gsm+07} analyzed both
radio and X-ray data taken in 2004 (day $\sim$6300) using a consistent method
and showed that the SNR sizes agree to within 1\%. Our results confirm their
finding: $R_1$ from the ring model at this epoch is fully consistent with the
X-ray radius reported by \citet[see Figure~(\ref{fig:fit}b)]{hbd+13}.

More generally, the ring fits allow a meaningful comparison between the radio
and X-ray SNR radii. From Figure~(\ref{fig:fit}b), the radius of the radio
remnant has exceeded that of the X-ray counterpart since day $\sim$7500. This
is also supported by the torus fits: \citet{ngm+09} employed the same torus
model as ours to fit the X-ray data taken on 2008 April (day 7736), and found
that it provides a slightly better fit than a simple ring, with a radius
of $0\farcs82\pm0\farcs02$ and $\theta=26\arcdeg\pm3\arcdeg$. Both values
are smaller than what we have obtained for the radio remnant at the same epoch
(0\farcs89\arcsec\ and 36\fdg2, respectively; Table~\ref{table:torus}),
suggesting a smaller extent of the remnant in X-rays than in radio. This 
agrees with the simulation results \citep{jn96}. In addition, the radio
emission may partly originate from fast shocks at high latitudes that have not
yet decelerated. This scenario can help explain the fact that the radio
remnant has apparently expanded beyond the optical inner ring with radius
$\sim0\farcs85$ \citep{plc+95}.

Figure~(\ref{fig:fit}c) shows that $\theta$ started to decrease at day
$\sim$7000, as did the ratio $R_2/R_1$ of the ring model. The results suggest
that the radio emission from lower latitudes has gradually dominated, which
could be a consequence of shock interaction with the dense circumstellar
medium in the equatorial ring. If the contribution of radio emission from high
latitudes continues to diminish, then we would expect the radius estimates
from both the torus and ring models to converge eventually.

The morphological evolution of SNR 1987A was accompanied by a reduction in the
degree of surface brightness asymmetry. Since the radio remnant first emerged,
the eastern lobe has always been brighter than the western one (see
Figure~\ref{fig:super}), and this has been attributed to faster shocks in the
east (e.g., \citealp{gms+97}; \citetalias{ngs+08}). This scenario is supported
by the significantly higher expansion velocities of the eastern lobe measured
from radio observations at higher frequencies \citep{zsn+13}, and by the
higher shock temperature found in the east from X-ray studies \citep{zmd+09}.
The faster shock in the east is expected to encounter the equatorial ring,
slow down and exit the ring earlier than in the west. This would reduce the
radio emissivity in the eastern rim and hence the overall brightness
asymmetry. If the observed trend continues, the western hemisphere of the
radio remnant may become brighter than the eastern hemisphere in a few years,
as predicted by 3D simulations of the expanding remnant (T. M. Potter et al.\
2013 in preparation). The same picture could also be applied to the X-ray
emission, which exhibits similar variations in the projected brightness
distribution \citep{nps+11b}.

In X-rays, \citet{hbd+13} reported drastic deceleration of the SNR from
8500\kms\ to 1820\kms\ on day $\sim$5900. However, we do not find conclusive
evidence for a similar deceleration in the radio emission. In particular,
$R_1$ from the ring fit shows a constant expansion at 3890\kms. For the torus
fit, while Figure~(\ref{fig:fit}b) suggests a break in the expansion, the
coupling between the radius and $\theta$ described above makes the
interpretation difficult. If the apparent break is physical, it would well
match the deceleration of the reverse shock predicted by one-dimensional
hydrodynamic simulations \citep[see Figure~9 of][]{ddh+12}. However, we
believe that the break is likely caused by the decreasing trend in $\theta$.
Hence, it reflects a change of the emission geometry rather than the slowing
down of the shock. Observing a deceleration of $R_1$ in future will confirm
this picture. This also gives a prediction that once $\theta$ stops shrinking,
the torus radius expansion rate should return to the same value as that
measured for $R_1$.

\section{Conclusion}
We have studied the evolution of the radio remnant of SN 1987A using ATCA
9\,GHz imaging observations taken between 1992 and 2013, and have carried out
Fourier modeling on the visibility data to quantitatively measure the remnant
structure. A truncated-shell torus model and an elliptical ring model were
used to fit the remnant morphology. They both suggest a gradual decrease in
the latitude extent of the remnant starting from day $\sim$7000, implying that
the radio emission from the equatorial region has progressively dominated.
This has been accompanied by a decreasing trend in the brightness asymmetry in
the east-west direction. Together these could indicate a new stage of the
remnant evolution, such that the forward shock has fully engulfed the entire
inner ring and is now interacting with the densest part of the circumstellar
medium. 

As a direct comparison between the torus and ring model fits, they give similar
results for most parameters, but the former always suggests a larger radius. The
discrepancy could be attributed to the projection effect or emission at high
latitudes to which the torus model is more sensitive. This also leads to
different expansion measurements. While the torus fit shows a break in the
expansion around day 7000 with the velocity slowing down from 4600\kms\ to
2400\kms, the ring fit indicates a constant expansion rate of 3890\kms. We
argue that the apparent break could be the result of coupling between the
torus radius and opening angle. We expect in the future when the latter stays
constant, both the torus and ring fits should give consistent expansion
velocity. Further observations at higher resolution (with VLBI or ALMA for
example) would be useful in understanding the true 3D nature of the evolving
remnant, and the time and latitude dependence of the expansion velocity.

\acknowledgements
We thank the referee for useful suggestions.
The Australia Telescope Compact Array is part of the Australia Telescope,
which is funded by the Commonwealth of Australia for operation as a National
Facility managed by CSIRO. Parts of this research were conducted by the
Australian Research Council Centre of Excellence for All-sky Astrophysics
(CAASTRO), through project number CE110001020.

{\it Facility:} \facility{ATCA}


\begin{thebibliography}{}
\expandafter\ifx\csname natexlab\endcsname\relax\def\natexlab#1{#1}\fi

\bibitem[{{Blondin} {et~al.}(1996){Blondin}, {Lundqvist}, \&
  {Chevalier}}]{blc96}
{Blondin}, J.~M., {Lundqvist}, P., \& {Chevalier}, R.~A. 1996, \apj, 472, 257

\bibitem[{{Burrows} {et~al.}(1995){Burrows}, {Krist}, {Hester}, {Sahai},
  {Trauger}, {Stapelfeldt}, {Gallagher}, {Ballester}, {Casertano}, {Clarke},
  {Crisp}, {Evans}, {Griffiths}, {Hoessel}, {Holtzman}, {Mould}, {Scowen},
  {Watson}, \& {Westphal}}]{bkh+95}
{Burrows}, C.~J., {Krist}, J., {Hester}, J.~J., {et~al.} 1995, \apj, 452, 680

\bibitem[{{Dewey} {et~al.}(2012){Dewey}, {Dwarkadas}, {Haberl}, {Sturm}, \&
  {Canizares}}]{ddh+12}
{Dewey}, D., {Dwarkadas}, V.~V., {Haberl}, F., {Sturm}, R., \& {Canizares},
  C.~R. 2012, \apj, 752, 103

\bibitem[{{Efron} \& {Tibshirani}(1993)}]{et93}
{Efron}, B., \& {Tibshirani}, R.~J. 1993, {An Introduction to the Bootstrap}
  (New York: Chapman and Hall)

\bibitem[{{Gaensler} {et~al.}(1997){Gaensler}, {Manchester}, {Staveley-Smith},
  {Tzioumis}, {Reynolds}, \& {Kesteven}}]{gms+97}
{Gaensler}, B.~M., {Manchester}, R.~N., {Staveley-Smith}, L., {et~al.} 1997,
  \apj, 479, 845

\bibitem[{{Gaensler} {et~al.}(2007){Gaensler}, {Staveley-Smith}, {Manchester},
  {Kesteven}, {Ball}, \& {Tzioumis}}]{gsm+07}
{Gaensler}, B.~M., {Staveley-Smith}, L., {Manchester}, R.~N., {et~al.} 2007, in
  AIP Conf. Proc., 937, Supernova 1987A: 20 Years After: Supernovae and
  Gamma-Ray Bursters, ed. S.~{Immler}, K.~{Weiler}, \& R.~{McCray} (Melville,
  NY: AIP), 86

\bibitem[{{Gull} \& {Daniell}(1978)}]{gd78}
{Gull}, S.~F., \& {Daniell}, G.~J. 1978, \nat, 272, 686

\bibitem[{{Helder} {et~al.}(2013){Helder}, {Broos}, {Dewey}, {Dwek}, {McCray},
  {Park}, {Racusin}, {Zhekov}, \& {Burrows}}]{hbd+13}
{Helder}, E.~A., {Broos}, P.~S., {Dewey}, D., {et~al.} 2013, \apj, 764, 11

\bibitem[{{Immler} {et~al.}(2007){Immler}, {Weiler}, \& {McCray}}]{iwm07}
{Immler}, S., {Weiler}, K., \& {McCray}, R., (eds.) 2007, AIP Conf. Proc., 
  937, {Supernova 1987A: 20 Years After: Supernovae and Gamma-Ray Bursters}
  (Melville, NY: AIP)

\bibitem[{{Jun} \& {Norman}(1996)}]{jn96}
{Jun}, B.-I., \& {Norman}, M.~L. 1996, \apj, 465, 800

\bibitem[{{Kemball} \& {Martinsek}(2005)}]{km05b}
{Kemball}, A., \& {Martinsek}, A. 2005, \aj, 129, 1760

\bibitem[{{Laki{\'c}evi{\'c}} {et~al.}(2012){Laki{\'c}evi{\'c}}, {Zanardo},
  {van Loon}, {Staveley-Smith}, {Potter}, {Ng}, \& {Gaensler}}]{lzv+12}
{Laki{\'c}evi{\'c}}, M., {Zanardo}, G., {van Loon}, J.~T., {et~al.} 2012, \aap,
  541, L2

\bibitem[{{Luo} {et~al.}(1994){Luo}, {McCray}, \& {Slavin}}]{lms94}
{Luo}, D., {McCray}, R., \& {Slavin}, J. 1994, \apj, 430, 264

\bibitem[{{Manchester} {et~al.}(2005){Manchester}, {Gaensler},
  {Staveley-Smith}, {Kesteven}, \& {Tzioumis}}]{mgs+05}
{Manchester}, R.~N., {Gaensler}, B.~M., {Staveley-Smith}, L., {Kesteven},
  M.~J., \& {Tzioumis}, A.~K. 2005, \apjl, 628, L131

\bibitem[{{Manchester} {et~al.}(2002){Manchester}, {Gaensler}, {Wheaton},
  {Staveley-Smith}, {Tzioumis}, {Bizunok}, {Kesteven}, \& {Reynolds}}]{mgw+02}
{Manchester}, R.~N., {Gaensler}, B.~M., {Wheaton}, V.~C., {et~al.} 2002, \pasa,
  19, 207

\bibitem[{{Morris} \& {Podsiadlowski}(2007)}]{mp07}
{Morris}, T., \& {Podsiadlowski}, P. 2007, Science, 315, 1103

\bibitem[{{Ng} {et~al.}(2009){Ng}, {Gaensler}, {Murray}, {Slane}, {Park},
  {Staveley-Smith}, {Manchester}, \& {Burrows}}]{ngm+09}
{Ng}, C.-Y., {Gaensler}, B.~M., {Murray}, S.~S., {et~al.} 2009, \apjl, 706,
  L100

\bibitem[{{Ng} {et~al.}(2008){Ng}, {Gaensler}, {Staveley-Smith}, {Manchester},
  {Kesteven}, {Ball}, \& {Tzioumis}}]{ngs+08}
{Ng}, C.-Y., {Gaensler}, B.~M., {Staveley-Smith}, L., {et~al.} 2008, \apj, 684,
  481 (N08)

\bibitem[{{Ng} {et~al.}(2011{\natexlab{a}}){Ng}, {Potter}, {Staveley-Smith},
  {Tingay}, {Gaensler}, {Phillips}, {Tzioumis}, \& {Zanardo}}]{nps+11}
{Ng}, C.-Y., {Potter}, T.~M., {Staveley-Smith}, L., {et~al.}
  2011{\natexlab{a}}, \apjl, 728, L15

\bibitem[{{Ng} {et~al.}(2011{\natexlab{b}}){Ng}, {Potter}, {Staveley-Smith},
  {Gaensler}, {Murray}, {Tingay}, {Phillips}, {Tzioumis}, \&
  {Zanardo}}]{nps+11b}
{Ng}, C.-Y., {Potter}, T.~M., {Staveley-Smith}, L., {et~al.}
  2011{\natexlab{b}}, \baas, 43, 20.11

\bibitem[{{Panagia}(1999)}]{pan99}
{Panagia}, N. 1999, in IAU Symp., 190, New Views of the Magellanic
  Clouds, ed. Y.-H. {Chu}, N.~{Suntzeff}, J.~{Hesser}, \& D.~{Bohlender},
 (Cambridge: Cambridge Univ.\ Press), 549

\bibitem[{{Plait} {et~al.}(1995){Plait}, {Lundqvist}, {Chevalier}, \&
  {Kirshner}}]{plc+95}
{Plait}, P.~C., {Lundqvist}, P., {Chevalier}, R.~A., \& {Kirshner}, R.~P. 1995,
  \apj, 439, 730

\bibitem[{{Potter} {et~al.}(2009){Potter}, {Staveley-Smith}, {Ng}, {Ball},
  {Gaensler}, {Kesteven}, {Manchester}, {Tzioumis}, \& {Zanardo}}]{psn+09}
{Potter}, T.~M., {Staveley-Smith}, L., {Ng}, C.-Y., {et~al.} 2009, \apj, 705,
  261

\bibitem[{{Racusin} {et~al.}(2009){Racusin}, {Park}, {Zhekov}, {Burrows},
  {Garmire}, \& {McCray}}]{rpz+09}
{Racusin}, J.~L., {Park}, S., {Zhekov}, S., {et~al.} 2009, \apj, 703, 1752

\bibitem[{{Sault} {et~al.}(1995){Sault}, {Teuben}, \& {Wright}}]{stw95}
{Sault}, R.~J., {Teuben}, P.~J., \& {Wright}, M.~C.~H. 1995, in ASP Conf. Ser.,
  Vol.~77, Astronomical Data Analysis Software and Systems IV, ed. R.~A.
  {Shaw}, H.~E. {Payne}, \& J.~J.~E. {Hayes} (San Francisco, CA: ASP), 433

\bibitem[{{Staveley-Smith} {et~al.}(1993{\natexlab{a}}){Staveley-Smith},
  {Briggs}, {Rowe}, {Manchester}, {Reynolds}, {Tzioumis}, \&
  {Kesteven}}]{sbr+93}
{Staveley-Smith}, L., {Briggs}, D.~S., {Rowe}, A.~C.~H., {et~al.}
  1993{\natexlab{a}}, \nat, 366, 136

\bibitem[{{Staveley-Smith} {et~al.}(2007){Staveley-Smith}, {Gaensler},
  {Manchester}, {Ball}, {Kesteven}, \& {Tzioumis}}]{sgm+07}
{Staveley-Smith}, L., {Gaensler}, B.~M., {Manchester}, R.~N., {et~al.} 2007, in
  AIP Conf. Proc., Vol. 937, Supernova 1987A: 20 Years After: Supernovae and
  Gamma-Ray Bursters, ed. S.~{Immler}, K.~{Weiler}, \& R.~{McCray} (Melville,
  NY: AIP), 96

\bibitem[{{Staveley-Smith} {et~al.}(1993{\natexlab{b}}){Staveley-Smith},
  {Manchester}, {Kesteven}, {Tzioumis}, \& {Reynolds}}]{smk+93}
{Staveley-Smith}, L., {Manchester}, R.~N., {Kesteven}, M.~J., {Tzioumis},
  A.~K., \& {Reynolds}, J.~E.~R. 1993{\natexlab{b}}, PASAu, 10, 331

\bibitem[{{Staveley-Smith} {et~al.}(1992){Staveley-Smith}, {Manchester},
  {Kesteven}, {Reynolds}, {Tzioumis}, {Killeen}, {Jauncey}, {Campbell-Wilson},
  {Crawford}, \& {Turtle}}]{smk+92}
{Staveley-Smith}, L., {Manchester}, R.~N., {Kesteven}, M.~J., {et~al.} 1992,
  \nat, 355, 147

\bibitem[{{Sugerman} {et~al.}(2005){Sugerman}, {Crotts}, {Kunkel}, {Heathcote},
  \& {Lawrence}}]{sck+05}
{Sugerman}, B.~E.~K., {Crotts}, A.~P.~S., {Kunkel}, W.~E., {Heathcote}, S.~R.,
  \& {Lawrence}, S.~S. 2005, \apjs, 159, 60

\bibitem[{{Suzuki} {et~al.}(1993){Suzuki}, {Shigeyama}, \& {Nomoto}}]{ssn93}
{Suzuki}, T., {Shigeyama}, T., \& {Nomoto}, K. 1993, \aap, 274, 883

\bibitem[{{Tingay} {et~al.}(2009){Tingay}, {Phillips}, {Amy}, {Tzioumis},
  {Kettenis}, {Boven}, {Szomoru}, {Paragi}, {van Langevelde}, {Verkouter},
  {Phillips}, {Cowie}, {Tam}, \& {Huisman}}]{tpa+09}
{Tingay}, S., {Phillips}, C., {Amy}, S., {et~al.} 2009, in 8th International
  e-VLBI Workshop, Proceedings of Science, 100

\bibitem[{{Turtle} {et~al.}(1990){Turtle}, {Campbell-Wilson}, {Manchester},
  {Staveley-Smith}, \& {Kesteven}}]{tcm+90}
{Turtle}, A.~J., {Campbell-Wilson}, D., {Manchester}, R.~N., {Staveley-Smith},
  L., \& {Kesteven}, M.~J. 1990, \iaucirc, 5086, 2

\bibitem[{{Wilson} {et~al.}(2011){Wilson}, {Ferris}, {Axtens}, {Brown},
  {Davis}, {Hampson}, {Leach}, {Roberts}, {Saunders}, {Koribalski}, {Caswell},
  {Lenc}, {Stevens}, {Voronkov}, {Wieringa}, {Brooks}, {Edwards}, {Ekers},
  {Emonts}, {Hindson}, {Johnston}, {Maddison}, {Mahony}, {Malu}, {Massardi},
  {Mao}, {McConnell}, {Norris}, {Schnitzeler}, {Subrahmanyan}, {Urquhart},
  {Thompson}, \& {Wark}}]{wfa+11}
{Wilson}, W.~E., {Ferris}, R.~H., {Axtens}, P., {et~al.} 2011, \mnras, 416, 832

\bibitem[{{Zanardo} {et~al.}(2013){Zanardo}, {Staveley-Smith}, {Ng},
  {Gaensler}, {Potter}, {Manchester}, \& {Tzioumis}}]{zsn+13}
{Zanardo}, G., {Staveley-Smith}, L., {Ng}, C.-Y., {et~al.} 2013, \apj, 767, 98

\bibitem[{{Zanardo} {et~al.}(2010){Zanardo}, {Staveley-Smith}, {Ball},
  {Gaensler}, {Kesteven}, {Manchester}, {Ng}, {Tzioumis}, \& {Potter}}]{zsb+10}
{Zanardo}, G., {Staveley-Smith}, L., {Ball}, L., {et~al.} 2010, \apj, 710, 1515

\bibitem[{{Zhekov} {et~al.}(2009){Zhekov}, {McCray}, {Dewey}, {Canizares},
  {Borkowski}, {Burrows}, \& {Park}}]{zmd+09}
{Zhekov}, S.~A., {McCray}, R., {Dewey}, D., {et~al.} 2009, \apj, 692, 1190

\end{thebibliography}

\end{document}